\documentstyle[eqsecnum,aps,prb,multicol,epsf]{revtex}
\begin{document}

\tighten

\title{Entanglement in Dilute Flux Line Liquids}
\author{A. M. Ettouhami}
\address{Department of Physics, University of Colorado, Boulder, CO 80309}
\date{\today}
\maketitle

\begin{abstract}
We construct, within the framework of classical statistical mechanics, a mean field theory of dilute flux line 
liquids which goes beyond linear hydrodynamics. Within our approach,
we find that interactions between vortices produce a massive term 
in the Hamiltonian of the internal modes of the flux lines which confines their transverse fluctuations.
This suggests that the flux line liquid, at least in the low density limit considered in this paper, might 
very well be in a weakly entangled state, where the average width 
$\langle u^2\rangle^{\frac{1}{2}}$ of flux lines can 
be much larger than the average distance between the lines but does not diverge 
with the sample thickness $L$. Consequences on the physics of flux line liquids are briefly discussed.
\end{abstract}

\pacs{64.60-i, 74.20.De, 74.60-w}

\begin{multicols}{2}

\section{Introduction}
\label{Intro}

The discovery, by Bednorz and M\"uller\cite{Bednorz-Muller} in 1986, of the new family of 
cuprate-based high temperature superconductors (HTSC), has had a profound impact on research in the field of 
superconductivity. 
In particular, due to their anisotropy, short coherence lengths and high critical temperatures, 
it quickly became obvious that thermal fluctuations will play a key role in the physics of interacting flux lines 
in these materials. Early experiments\cite{Gammel1,Gammel2,Safar1,Safar2,Charalambous} 
suggested, and it was later unambiguously confirmed\cite{Kwok1,Kwok2,Cubitt,Zeldov}, that a phase transition takes 
place where the three dimensional Abrikosov\cite{Abrikosov} flux line lattice (FLL) 
lattice melts into a flux line liquid, and the 
location of the melting temperature $T_m$ as a function of the applied magnetic field $H$ in the $(H,T)$ 
phase diagram of the flux line system became the subject of 
intense experimental and 
theoretical\cite{Nelson,Nelson-Seung,Houghton-et-al,Ma-Chui,Sengupta,Li-Teitel,Ryu-et-al,Hetzel-et-al,Dodgson-et-al} 
investigations (see Blatter et {\em al.}\cite{Blatter-et-al} for a recent review).

One of the early approaches aimed at describing the melting transition and the physics of melted flux line 
liquids was the theory of Nelson and Seung\cite{Nelson-Seung}, which is based on the 
observation\cite{Fisher-Lee,Nelson} that the partition function of a system of interacting flux lines in
$2+1$ dimensions can be mapped onto the imaginary time partition function of a system of interacting quantum 
mechanical bosons in two dimensions. 
This formal mapping was used to obtain a complete description of flux line liquids by using
well known techniques from boson physics.
The physical picture of the flux line liquid which emerges from this description suggests that 
there is considerable wandering of flux lines as they traverse the 
sample and that the flux line liquid is heavily entangled over much of the liquid phase.
Such a picture has been observed in numerical simulations, and it is now generally believed that
the mean square relative displacement $\langle[{\bf r}_i(z)-{\bf r}_i(0)]^2\rangle$
of the transverse position ${\bf r}_i(z)$ of a given flux line in a flux 
line liquid has the same form as the corresponding quantity for an isolated flux line, namely~:
\begin{eqnarray}
\langle|{\bf r}_i(z) - {\bf r}_i(0)|^2\rangle = 2 D|z|  \qquad ,
\label{correlator}
\end{eqnarray}
the only effect of the interactions between lines in a dense liquid being to reduce\cite{Marchetti} the 
``diffusion'' constant $D$ from its bare value $D_0$ for a single flux line, which is given 
by\cite{Nelson,Nelson-Seung}~:
\begin{eqnarray}
D_0=\frac{k_BT}{\tilde\varepsilon_1}
\nonumber
\end{eqnarray}
where $k_B$ is Boltzmann's constant, $T$ is the temperature and $\tilde\varepsilon_1$ is the tilt modulus of a 
single flux line. Similarly, the mean square projected area occupied by the lines 
is found, within the boson picture, to be of order\cite{Nelson-Seung,Nelson2}
\begin{eqnarray}
\langle u^2\rangle \simeq \frac{2\pi k_BT}{\tilde\varepsilon_1}\,L \label{area}
\end{eqnarray}
where ${\bf u}(z)={\bf r}(z)-\langle{\bf r}(z)\rangle$ and
$L$ is the sample thickness in the direction of the flux lines (which we take parallel to the 
$\hat{\bf z}$ axis).
This last result is to be contrasted with the mean square projected area of a given flux line in the crystalline 
phase, which is given by\cite{Blatter-et-al}~:
\begin{eqnarray}
\langle u^2\rangle = \frac{k_BTa}{2\sqrt{\pi}\,\varepsilon\varepsilon_0}
\label{usqrd-crystal}
\end{eqnarray}
where $a$ is the average spacing between flux lines in the vortex lattice, $\varepsilon=\lambda/\lambda_c$ is 
the ratio of the London penetration depths in the $(ab)$ plane and along the direction of the ${\bf c}$ axis 
respectively (we remind the reader that most HTSC have uniaxial symmetry, and we denote by ${\bf c}$ the 
principal axis of symmetry of the crystal), and $\varepsilon_0=(\phi_0/4\pi\lambda)^2$, with $\phi_0=hc/2e$ 
the flux quantum\cite{deGennes}. 
Comparing the results (\ref{area}) and (\ref{usqrd-crystal}), we see that, while the mean square projected area of 
flux lines $\langle{u^2}\rangle$ in the crystalline phase is finite, in the liquid phase it diverges with the sample 
thickness $L$.

Despite a good deal of both theoretical 
\cite{Marchetti-Nelson1,Marchetti-Nelson2,Radzihovsky-Frey,Tesanovic1,Tesanovic2,Benetatos1,Benetatos2}
and experimental work\cite{Safar-et-al-prl-94,delaCruz-94,Lopez-et-al-prl-96,Righi-et-al} 
and rather extensive numerical studies
\cite{Li-Teitel-prl-91,Li-Teitel-prb-93,Carneiro-et-al-prb-93,Chen-Teitel-prl-94,Li-Teitel-prb-94,Carneiro-prb94,Chen-Teitel-prl-95,Carneiro-prl-95,Nguyen-Sudbo-Hetzel-prl-96,Carneiro-prb-96,Hagenaars-et-al,Chen-Teitel-prb-97-1,Chen-Teitel-prb-97-2,Koshelev,Ryu-Stroud-prl,Ryu-Stroud,Nordborg-Blatter-prl-97,Nordborg-Blatter-prb-98,Nguyen-Sudbo-prb-98-1,Nguyen-Sudbo-prb-99,Olsson-Teitel,Chin-et-al}, 
several questions regarding the physics of flux line liquids remain open. 
One of the most important
questions which has been the subject of intense debate during the past few 
years\cite{Feigelman-et-al,Moore}, 
is related to whether a disentangled flux line liquid can survive above melting.
Although early numerical simulations have provided support for the proposal by 
Feigel'man et {\em al.}\cite{Feigelman-et-al}
that superconducting coherence could survive in the liquid phase, more extensive simulations by 
several authors\cite{Hu-et-al,Nguyen-Sudbo-prb-98-2,Nordborg-Blatter-prb-98,Olsson-Teitel} 
found that longitudinal coherence vanishes simultaneously with melting, 
and that the vortex liquid is characterized by very short 
entanglement lengths in the direction of the flux lines.
However, analyses of experiments on high quality untwinned single crystals of 
YBa$_2$Cu$_3$O$_{7-\delta}$ (YBCO) by Righi et {\em al.}\cite{Righi-et-al} and by 
Moore\cite{Moore} have suggested that longitudinal correlations in the liquid phase 
may be of surprizingly larger micron scale.

At the fundamental level, it is very important to understand,
from a purely conceptual point of view, under what circumstances
a disentangled state can emerge from a theoretical description, and what kind of theories are best suited to yield 
such a disentangled state. We first note that, for a disentangled state to appear, it is necessary that flux line 
conformation variables ${\bf r}_i(z)$ have correlations which do {\em not} obey equations
(\ref{correlator})-(\ref{area}). In particular, the average projected area of a given flux line 
$\langle u^2\rangle$ has to be independent of the sample thickness $L$.
In the crystalline phase, transverse fluctuations of a given flux line are confined thanks to the repulsive action 
of the other, neighboring lines in the FLL.
In a similar way, physical intuition suggests that the interactions between flux lines should strongly
reduce\cite{Marchetti} and might even suppress line wandering if the repulsion energy between flux
lines is high enough compared to $k_BT$. In such a disentangled state with large repulsion forces between the 
lines, a given flux line will experience the repulsive potential
of its surrounding neighbors in much the same way as in a lattice. As a result, the internal fluctuations
of that line will be confined in about the same way they are confined in regular Abrikosov FLL.
In particular, there should be a regime where the mean
square internal fluctuations $\langle u^2\rangle$ does not differ much, at least locally,
from the corresponding quantity in a FLL at the same density.
For us to be able to describe a putative disentangled phase of a flux line liquid, 
that is, a phase where fluctuations of flux lines are confined, we need to concentrate on the individual 
conformation variables of the flux lines, and try to evaluate the effect of interactions on these variables in the 
most accurate way possible.

In fact, most theoretical studies to date have concentrated on hydrodynamic
quantities, such as, {\em e.g.} the density,
and have used approximations pertinent to linear hydrodynamics
in order to study the physics of flux line liquids. Hydrodynamic observables being coarse-grained quantities, such
approaches are intrinsically unable to give an accurate picture of the correlations of flux line
conformation variables, such as ${\bf u}_i(z)$. Even if we go
beyond linear hydrodynamics, such as in ref.\cite{Benetatos1}, 
because of the use of coarse-grained variables, we expect 
not to be able to describe the confinement of individual flux line fluctuations that would take place in a 
hypothetical, disentangled flux line liquid. 
To the most, hydrodynamic approaches can only yield a renormalization of the tilt modulus of 
flux lines in a line liquid, leaving the qualitative (analytic) form of the correlations 
(\ref{correlator})-(\ref{area}) unchanged.

In view of the above remarks, it seems to us that
it is highly desirable to construct a theory which keeps track in a better way of
conformation variables of the flux lines, and which would therefore be able to yield more accurate information 
about the correlations of these variables. Shuch a theory would undoubtedly help us
gain a better understanding of the properties of flux line liquids.

In this paper, which is largely motivated by the conclusions of reference\cite{Righi-et-al}, 
we would like to lay out the path for the construction of such a theory. 
We show, in particular, that a classical 
theory of flux liquids with a better handling of the internal modes of flux lines
can be formulated which is simple enough to allow for ease of mathematical treatment, and 
which is transparent enough to allow for approximations to be made with a good level of confidence.
Within this approach, and under {\em certain assumptions}, we find 
that a confinement of the internal fluctuations of the lines can indeed take place,
at least in the limit of a dilute flux line liquid. More specifically, we find 
that, due to the interactions between flux lines, the internal modes
may actually become massive, leading to a finite mean square width 
$\langle u^2\rangle^{\frac{1}{2}}$ of flux lines, and to a 
qualitative change in the behaviour of the correlation function
(\ref{correlator}).

This paper is organized as follows. 
In section \ref{Theory-disentangled}, we consider a liquid of interacting flux lines in a weakly anisotropic HTSC. 
Performing a perturbative expansion of the
interaction energy in terms of the internal fluctuations of the flux lines, we show that the interactions with
neighboring lines induce a mass term in the Hamiltonian of the internal modes. 
As a result, we find that the internal fluctuations of the flux lines are strongly suppressed compared with the
free flux line result (\ref{area}).
We self-consistently determine the range of validity of our perturbative expansion, which is found to be quite
large for moderately anisotropic superconductors. 
We also calculate the structure factor of the flux line liquid 
and our result is compared to previous derivations.
Section \ref{Conclusions} contains a discussion of our findings, which we think can be reconciled with the results 
of numerical simulations, along with our conclusions.

\section{Mean-field theory of (putative) disentangled flux line liquids}
\label{Theory-disentangled}

In this section, we shall be considering the statistical mechanics of an assembly of flux lines in 
a sample of thickness $L$ of a uniaxial 
HTSC, with both the principal axis of anisotropy ${\bf c}$ and the external magnetic field
aligned with the $\hat{\bf z}$ axis.
Such a system can be described by the Hamiltonian\cite{Nelson-Seung}
\begin{eqnarray}
H & = & \sum_{i=1}^N\int_0^L dz\,\,\frac{1}{2}\,\tilde\varepsilon_1\,\Big(\frac{d{\bf r}_i}{dz}\Big)^2 +
\nonumber\\
& + &\frac{1}{2}\sum_{i\neq j}\int_0^L dz\,\, V\big({\bf r}_i(z) - {\bf r}_j(z)\big)
\label{interH}
\end{eqnarray}
where $\tilde\varepsilon_1\simeq\varepsilon^2\varepsilon_0\ln(a/{\xi})$ denotes the tilt modulus 
per unit length of the flux
lines (here ${\xi}$ and $a$ are the coherence length and the average distance between flux lines 
in the $(ab)$ plane respectively), and
where $V({\bf r})=2\varepsilon_0K_0(r/{\lambda})$ is the interaction potential between flux line elements at
equal height, with $K_0$ a modified Bessel function\cite{deGennes,Abramowitz}.
(Henceforth, we shall neglect the logarithmic factor $\ln(a/{\xi})$, of order unity, in the definition of the 
tilt modulus, writing $\tilde\varepsilon_1\approx\varepsilon^2\varepsilon_0$.)
The Hamiltonian above may be viewed as a simplified version of the most general Hamiltonian of 
arbitrarily curved and tilted flux lines in the London regime of anisotropic superconductors  
\cite{Brandt,Barford-Gunn,Sudbo-Brandt,Sardella} (with nonlocal 
interactions in the $z$ direction), and can be shown to be a good 
approximation to the latter in the limit of nearly straight flux lines\cite{remark0}. 
The tilt modulus per unit {\em length}
$\tilde\varepsilon_1$ on the other hand can be viewed as the effective tilt modulus for long-wavelength $q_z$ 
distortions which accounts for the most relevant fluctuations of the flux lines near melting, with 
transverse wavelength $q_\perp\sim q_{BZ}$ ($q_{BZ}$ being the transverse wavevector at the Brillouin zone 
boundary), as can be seen by replacing $q_\perp\simeq q_{BZ}=(4\pi B/\phi_0)^{1/2}$, 
taking the limit $q_z\to 0$ in the general (nonlocal) expression of the tilt modulus per unit {\em volume} 
$c_{44}({\bf q})=B^2/4\pi(1+(\lambda/\varepsilon)^2q_\perp^2+\lambda^2q_z^2)$,
and multiplying by the elementary area per vortex $(\phi_0/B)$.
Below, it will prove useful to write the following 
Fourier decomposition
of ${\bf r}_i(z)$ into Rouse modes\cite{Doi-Edwards}~:
\begin{eqnarray}
{\bf r}_i(z) = \sum_{n=-\infty}^\infty {\bf r}_i(q_n)\,\mbox{e}^{iq_nz}
\end{eqnarray}
where $q_n=2\pi{n}/L$, and where the coefficients ${\bf r}_i(q_n)$ are given by~:
\begin{eqnarray}
{\bf r}_i(q_n) = \frac{1}{L}\int_0^L dz\; {\bf r}_i(z)\,\mbox{e}^{-iq_nz}
\end{eqnarray}
as can be verified by using the orthogonality relation
\begin{eqnarray}
\int_{0}^L dz \;\mbox{e}^{iq_nz}(\mbox{e}^{iq_mz})^* = L\delta_{n,m}
\label{orthog-rel}
\end{eqnarray}
It will also be convenient to write ${\bf r}_i(z)$ as the sum
\begin{eqnarray}
{\bf r}_i(z) = {\bf r}_{0i} + {\bf u}_i(z)
\label{decomp}
\end{eqnarray}
where 
\begin{eqnarray}
{\bf r}_{0i} = {\bf r}_i(q_n=0) = \frac{1}{L}\int_0^L dz\; {\bf r}_i(z) \nonumber
\end{eqnarray}
is the center of mass (CM) position, while
${\bf u}_i(z)$ is the displacement of the flux line at height $z$ 
with respect to the center of mass position. 
In terms of the Fourier modes $\{{\bf r}_i(q_n)\}$, the displacement vector ${\bf u}_i(z)$ 
can be written in the form (here $c.c.$ denotes complex conjugation)~:
\begin{eqnarray}
{\bf u}_i(z) = \sum_{n=1}^\infty\{ {\bf r}_i(q_n)\mbox{e}^{iq_nz} + c.c. \}
\label{fourier-internal}
\end{eqnarray}

Using the above decomposition of the flux line positions $\{{\bf r}_i(z)\}$ 
into CM and internal modes, equation (\ref{decomp}),
the interaction term between flux line elements at ${\bf r}_i(z)$ and ${\bf r}_j(z)$ can be written as
\begin{eqnarray}
V\big({\bf r}_i(z)-{\bf r}_j(z)\big) = V\big({\bf r}_{0i}-{\bf r}_{0j} + {\bf u}_i(z)-{\bf u}_j(z)\big)
\label{interm1}
\end{eqnarray}
Since the interaction potential $V(r)$ is a smooth function which 
varies very slowly on length scales much smaller than ${\lambda}$, we
see that if 
\begin{equation}
|{\bf u}_i(z)-{\bf u}_j(z)| \ll {\lambda}
\label{condition}
\end{equation}
then we can write, to a very good approximation~:
\begin{eqnarray}
V\big({\bf r}_i(z) &-& {\bf r}_j(z)\big) \simeq  V\big({\bf r}_{0i}-{\bf r}_{0j}\big) + 
\nonumber\\
& + & [{\bf u}_i(z)-{\bf u}_j(z)]\cdot
\nabla V\big({\bf r}_{0i}-{\bf r}_{0j}\big) +\nonumber\\
& + & \frac{1}{2}\big([{\bf u}_i(z)-{\bf u}_j(z)]\cdot\nabla\big)^2V\big({\bf r}_{0i}-{\bf r}_{0j}\big)
\label{Taylorexpansion}
\end{eqnarray}
If we integrate this last equation over $z$,
taking into account the fact that $\int_0^L dz\; {\bf u}_i(z) = 0$, we obtain
\begin{eqnarray}
\int_0^L &{dz}& \,\,V\big({\bf r}_i(z) - {\bf r}_j(z)\big) \simeq LV\big({\bf r}_{0i}-{\bf r}_{0j}\big) +
\nonumber\\
&+&\frac{1}{2}\int_0^L dz\,\,
\big[\, u_{i\alpha}(z)u_{i\beta}(z) + u_{j\alpha}(z)u_{j\beta}(z) + \nonumber\\
&-& u_{i\alpha}(z)u_{j\beta}(z)-u_{j\alpha}(z)u_{i\beta}(z)\big]\,
\partial_\alpha\partial_\beta V\big({\bf r}_{0i}-{\bf r}_{0j}\big) \nonumber
\end{eqnarray}
In the Boson language, this last equation is reminiscent of the expansion of the action of a quantum particle 
around a classical path \cite{Feynman-Hibbs}. 
The Hamiltonian (\ref{interH}) now can be written in the form~:
\begin{eqnarray}
H = H_0 + H_1
\end{eqnarray}
where
\begin{eqnarray}
H_0 = \frac{1}{2}\sum_{i\neq j} LV({\bf r}_{0i}-{\bf r}_{0j})
\label{H0}
\end{eqnarray}
is the Hamiltonian of a system of perfectly straight, interacting lines of length $L$, and 
\begin{eqnarray}
H_1 & = & 
\sum_{i=1}^N\int_0^L dz\,\Big[\,\frac{1}{2}\,\tilde\varepsilon_1\,\Big(\frac{d{\bf u}_i}{dz}\Big)^2 +
\frac{1}{2}\;\mu^{(i)}_{\alpha\beta}u_{i\alpha}(z)\,u_{i\beta}(z)\,\Big] +
\nonumber\\ 
& + & \frac{1}{2}\sum_{i=1}^N\sum_{j\neq i}\int_0^L dz\,\,\mu^{(ij)}_{\alpha\beta}u_{i\alpha}(z)\,u_{j\beta}(z)
\label{TaylorHam1}
\end{eqnarray}
is the Hamiltonian of the internal modes of the flux lines. In equation (\ref{TaylorHam1}), we defined the
following quantities~:
\begin{eqnarray}
\mu^{(i)}_{\alpha\beta} & = & \sum_{j(\neq i)=1}^N\partial_\alpha\partial_\beta
V\big({\bf r}_{0i}-{\bf r}_{0j}\big) \label{mu(i)}
\\
\mu^{(ij)}_{\alpha\beta} & = & -\partial_\alpha\partial_\beta
V\big({\bf r}_{0i}-{\bf r}_{0j}\big) \label{mu(ij)}
\end{eqnarray}

As can be seen from equation (\ref{TaylorHam1}), the interactions between flux lines 
have generated a mass term $\frac{1}{2}\mu^{(i)}_{\alpha\beta}u_{i\alpha}(z)u_{i\beta}(z)$ 
for the internal modes $\{{\bf u}_i(z)\}$ as well as 
additional, two-body terms, $\frac{1}{2}\mu^{(ij)}_{\alpha\beta}u_{i\alpha}(z)u_{j\beta}(z)$,
which couple the internal modes of different vortices $(i\neq j)$. 
As they stand, the coefficients
$\mu^{(i)}_{\alpha\beta}$ and $\mu^{(ij)}_{\alpha\beta}$ of equations (\ref{mu(i)}) and (\ref{mu(ij)})
depend on the specific configuration of all the center of mass coordinates $\{{\bf r}_{0i}\}$ considered. 
In order to obtain the value of these coefficients relevant to a liquid of flux lines, 
we need to average equations (\ref{mu(i)}) and (\ref{mu(ij)}) over all possible configurations of the CM 
coordinates compatible with a liquid structure. 
In order to carry out such an average, let us write the partition
function of the flux line liquid ($\beta=1/k_BT$ is the inverse temperature) in the form
\begin{eqnarray}
{\cal Z} = \mbox{Tr}_0\,\mbox{Tr}_u\big(\mbox{e}^{-\beta H}\big) \label{partfunc}
\end{eqnarray}
where we denote by $\mbox{Tr}_0$ and $\mbox{Tr}_u$ the trace over the center of mass and internal modes, 
respectively. More explicitely~:
\begin{eqnarray}
\mbox{Tr}_0 & = & \int d{\bf r}_{01}\cdots d{\bf r}_{0N} \\
\mbox{Tr}_u & = & \int\prod_{n\ge 1} d{\bf r}_1(q_n)\cdots \int\prod_{n\ge 1} d{\bf r}_N(q_n)
\end{eqnarray}
where $d{\bf r}_i(q_n)$ stands for $d{\bf r}_{i,re}(q_n)\,d{\bf r}_{i,im}(q_n)$;
${\bf r}_{i,re}(q_n)$ and ${\bf r}_{i,im}(q_n)$ being the real and imaginary parts of ${\bf r}_i(q_n)$, 
respectively. Noting that $H=H_0+H_1$ in (\ref{partfunc}), and taking the trace over the CM mode only, 
we obtain
\begin{eqnarray}
\mbox{Tr}_0\big(\mbox{e}^{-\beta H}\big) = 
Z_0\big\langle\mbox{e}^{-\beta H_1}\big\rangle_0 \label{interm-cumul}
\end{eqnarray}
where $Z_0=\mbox{Tr}_0(\mbox{e}^{-\beta H_0})$, and where the average $\langle\cdots\rangle_0$ is taken with
respect to the probability distribution $\exp(-\beta H_0)/Z_0$, {\em i.e.}
for an arbitrary function $f(\{{\bf r}_{0i}\})$ of the CM coordinates 
$\{ {\bf r}_{01},\cdots,{\bf r}_{0N} \}$,
\begin{eqnarray}
\big\langle f(\{{\bf r}_{0i}\})\big\rangle_0 = 
\frac{1}{Z_0}\int d{\bf r}_1\dots d{\bf r}_N\,\, f(\{{\bf r}_{0i}\}) \,\mbox{e}^{-\beta H_0}
\label{avgliquid}
\end{eqnarray}
We now perform a cumulant expansion in equation (\ref{interm-cumul}), with the following result to leading order 
in $H_1$,
\begin{eqnarray}
\mbox{Tr}_0\big(\mbox{e}^{-\beta H}\big) \simeq \,Z_0\,\,\mbox{e}^{-\beta\langle H_1\rangle_0}
\end{eqnarray}
and hence we see that the total partition function ${\cal Z}$ can be written in the form
\begin{eqnarray}
{\cal Z} \simeq Z_0 Z_1
\end{eqnarray}
where $Z_1 = \mbox{Tr}_u(\mbox{e}^{-\beta\langle H_1\rangle_0})$ is the effective partition function of the
internal modes. The quantity
$$H_{eff} = \langle H_1\rangle_0$$
may be thought of as the effective Hamiltonian of the internal modes, 
averaged over all possible CM liquid configurations. 
In order to find this effective Hamiltonian, all we need to do is to calculate the averages
$\langle\mu^{(i)}_{\alpha\beta}\rangle_0$ and $\langle\mu^{(ij)}_{\alpha\beta}\rangle_0$.
Applying the definition of the average (\ref{avgliquid}) to the 
``mass'' coefficients $\mu^{(i)}_{\alpha\beta}$, we
obtain
\begin{eqnarray}
\langle\mu^{(i)}_{\alpha\beta}\rangle_0 & = &\frac{1}{Z_0}\int \prod_{k=1}^Nd{\bf r}_{0k}\,\,
\sum_{j(\neq i)=1}^N \partial_\alpha\partial_\beta V({\bf r}_{0i}-{\bf r}_{0j})\mbox{e}^{-\beta H_0} \nonumber\\
& = & \sum_{j(\neq i)=1}^N\frac{1}{Z_0}\int d{\bf r}_{0i}d{\bf r}_{0j}\,\,
\partial_\alpha\partial_\beta V({\bf r}_{0i}-{\bf r}_{0j})\times\nonumber\\
&\times&\int\prod_{k\neq i,j}d{\bf r}_{0k}\,\,\mbox{e}^{-\beta H_0}
\label{interm2}
\end{eqnarray}
The coordinates $\{{\bf r}_{0l}\}$ under the integral sign being just dummy variables,
we see that the {\em rhs} of equation (\ref{interm2}) consists of a sum of $(N-1)$ identical terms which
can be rewritten in the form
\begin{eqnarray}
\langle\mu^{(i)}_{\alpha\beta}\rangle_0 & = & \frac{(N-1)}{Z_0}\int d{\bf r}_{01}d{\bf r}_{02}\,\,
\partial_\alpha\partial_\beta V({\bf r}_{01}-{\bf r}_{02}) \times \nonumber\\
&\times&\int d{\bf r}_{03}\dots d{\bf r}_{0N}\;\;\mbox{e}^{-\beta H_0}
\label{interm3}
\nonumber\\
& = & \frac{1}{N}\int d{\bf r}_{01}d{\bf r}_{02}\,\,
\partial_\alpha\partial_\beta V({\bf r}_{01}-{\bf r}_{02})\times\nonumber\\
&\times&\frac{N(N-1)}{Z_0}\int d{\bf r}_{03}\dots d{\bf r}_{0N}\;\;\mbox{e}^{-\beta H_0}
\label{interm4}
\end{eqnarray}
We recognize on the second line of equation (\ref{interm4})
the two-particle density $\rho^{(2)}({\bf r}_1,{\bf r}_2)$ of liquid state
theory \cite{McQuarrie}
\begin{eqnarray}
\rho^{(2)}({\bf r}_{01},{\bf r}_{02}) = 
\frac{N(N-1)}{Z_0}\int d{\bf r}_{03}\dots d{\bf r}_{0N}\,\mbox{e}^{-\beta H_0}
\end{eqnarray}
This last quantity is related to the pair distribution function 
of the CM mode $g_0({\bf r}_{01},{\bf r}_{02})$ by
$$\rho^{(2)}({\bf r}_{01},{\bf r}_{02}) = \rho^2 g_0({\bf r}_{01},{\bf r}_{02})$$
where $\rho = B/\phi_0$ is the average density of flux lines in the system. 
At equilibrium, we expect
the flux line liquid to be translationally invariant, which implies that 
$g_0({\bf r}_{01},{\bf r}_{02})$ will depend only on 
the difference $({\bf r}_{01}-{\bf r}_{02})$, {\em i.e.} 
$g_0({\bf r}_{01},{\bf r}_{02})=g_0({\bf r}_{01}-{\bf r}_{02})$. Under these
conditions, we obtain from equation (\ref{interm4})
\begin{eqnarray}
\langle\mu^{(i)}_{\alpha\beta}\rangle_0 & = & \frac{\rho^2}{N}\int d{\bf r}_{01}d{\bf r}_{02}\;
g_0({\bf r}_{01}-{\bf r}_{02})\partial_\alpha\partial_\beta V({\bf r}_{01}-{\bf r}_{02}) \nonumber\\
& = & \rho \int d{\bf r}\,\, g_0({\bf r})\partial_\alpha\partial_\beta V({\bf r})
\label{result-mu-g}
\end{eqnarray}
Introducing the pair correlation function $h_0({\bf r}) = g_0({\bf r}) - 1$, 
the {\em rhs} of the last equation becomes 
\begin{eqnarray}
\langle\mu^{(i)}_{\alpha\beta}\rangle_0 = \rho \int d{\bf r}\; 
h_0({\bf r})\partial_\alpha\partial_\beta V({\bf r})
\end{eqnarray}
where we used the fact that $\int d{\bf r}\,\,\partial_\alpha\partial_\beta V({\bf r}) = 0$.
If we further use the fact that both $h_0$ and $V$ are rotationally invariant functions, 
$h_0({\bf r})=h_0(r)$ and $V({\bf r})=V(r)$ , we obtain (here $d_\perp=2$ is the number of transverse dimensions)
\begin{eqnarray}
\langle\mu^{(i)}_{\alpha\beta}\rangle_0 & = & \mu\,\delta_{\alpha,\beta} \label{avg-diag}\\
\mu & = & \,\frac{\rho}{d_\perp}\,\int d{\bf r}\; h({\bf r})\nabla_\perp^2V({\bf r})
\label{result1mu(i)}
\end{eqnarray}
where $\nabla_\perp^2=\partial_x^2+\partial_y^2$ is the Laplacian in the transverse directions.
Equation (\ref{result1mu(i)}) is the main result of this paper. It gives the effective mass of the internal
modes of flux lines, averaged over all possible configurations of the centers of mass of the vortices in the flux
line liquid.

From equation (\ref{result1mu(i)}) above, we now want to derive
an explicit expression for the averaged mass $\mu$. In order to be able to do so, we need to
choose a suitable analytic form for the pair correlation function $h_0(r)$. 
It is well known from liquid state theory\cite{McQuarrie,Hansen} 
that the pair distribution function $g(r)$ of an ordinary classical liquid 
has peaks of decreasing amplitude at $r\approx a,\, 2a,\,3a,\dots$; 
$a=1/\sqrt{\rho}$ being the average spacing between particle in the liquid phase;
and that it has the limiting behaviors $g(r)\ll 1$ for $r < a$, and $g(r)\to 1$ when $r\to\infty$. A common
(mean field) approximation for the pair distribution function used in the literature consists in neglecting 
the fine structure of
$g(r)$ for $r>a$, and approximating
\begin{eqnarray}
g(r) \simeq \theta(r-a)
\label{crude-g0}
\end{eqnarray}
where $\theta$ is the Heaviside unit step function ($\theta(x)=1$ if $x>0$ and zero otherwise).
This approximation, which is obviously relevant to liquids at not very high densities, retains the important
feature of $g(r)$ that it is very small for $r<a$, which expresses the fact
that particles in a liquid of density $\rho$ have a small probability of being within a distance  
$a=1/\sqrt\rho$ from each other. 
Here, we expect such a behaviour of the pair distribution function $g_0(r)$ in a dilute flux liquid to be valid if 
the repulsive interactions between flux lines are strong enough. To quantify how strong the interactions must be, 
we introduce the following typical energy scale 
\begin{equation}
E_c(T) = \varepsilon_0\; l_c(T)
\end{equation}
where $l_c(T)=\mbox{max}\big(s,\xi_c(T)\big)$, $s$ being the distance between superconducting planes in our 
layered material. $E_c$ represents the typical energy barrier for the crossing of two flux lines nearly 
parallel to ${\bf c}$, and if $E_c(T)\gg k_BT$ we expect interactions between lines to strongly reduce cutting and 
crossing of lines, and as a consequence, to

\begin{minipage}[t]{3.2in}
\vspace{0.1in}  
\epsfxsize=2.2in
\hfill \epsfbox{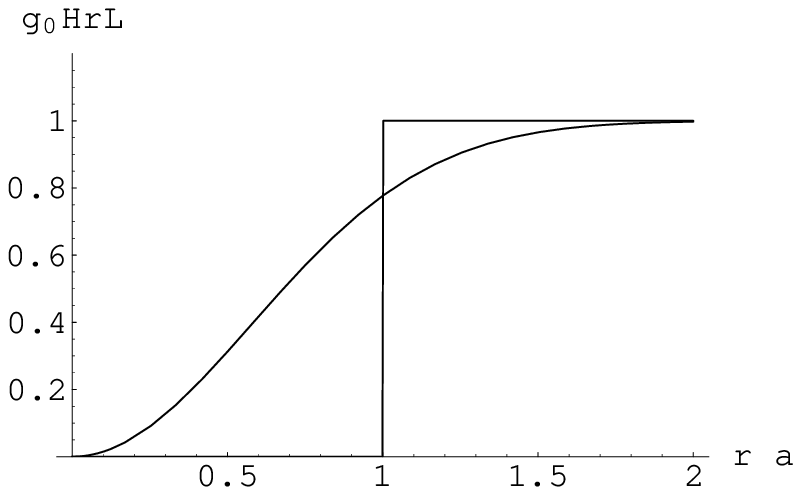} \hfill
\begin{small}

Figure 1. Pair distribution function $g_0(r)$ of equation (\ref{pair-distr-func}) for $\alpha = 3/2$.
Also shown is the mean-field approximation $g_0(r) =\theta(r-a)$.
\end{small}
\end{minipage}

\medskip

\noindent lead to the kind of confinement of the internal modes we find
above. In this regime, equation (\ref{crude-g0}) should be a good 
approximation to the pair distribution function $g_0(r)$ in a dilute flux liquid.
Although the condition $E_c\gg k_BT$ may prove to be too stringent for actual flux lines in a HTSC, 
we here consider this limit as an extreme case for which a disentangled state might be realized.

For the purpose of mathematical tractability, we shall here use the approximation (\ref{crude-g0}) in the form
(figure 1)
\begin{eqnarray}
g_0(r) = 1 - \exp(-\alpha r^2/a^2)
\label{pair-distr-func}
\end{eqnarray}
which gives for the pair correlation function
\begin{eqnarray}
h_0(r) = - \exp(-\alpha r^2/a^2) \label{pair-corr-fct}
\end{eqnarray}
The numerical coefficient $\alpha$ inside the exponential is to be chosen so that 
$|h_0(r=a)|\ll 1$. 
Using this form of the pair correlation function into equation (\ref{result1mu(i)}), we obtain~:
\begin{eqnarray}
\mu = \frac{\pi}{d_\perp\alpha}\,(\rho a^2)\int\frac{d^2{\bf k}}{(2\pi)^2}\,\, k^2V({\bf k})
\,\mbox{e}^{-k^2a^2/2\alpha} \label{equation-mu}
\end{eqnarray}
Inserting the Fourier transform of the interaction potential 
\begin{equation}
V({\bf k}) = \frac{4\pi\varepsilon_0}{k^2+{\lambda}^{-2}}
\label{interaction-k}
\end{equation}
we obtain, after a few manipulations (Appendix A)
\begin{eqnarray}
\mu \approx \frac{2\pi}{d_\perp}\;\rho\varepsilon_0 \label{resultmu}
\end{eqnarray}
where we assumed that $a<{\lambda}$. It is interesting to note that the final result is, to a very good
approximation, independent of the
particular choice of $\alpha$ in equation (\ref{pair-corr-fct}), as long as this last condition ($a<{\lambda}$)
is satisfied.
It is also interesting to note that a nontrivial structure in 
$g_0(r)$ is necessary to obtain a finite value for 
the ``mass'' $\mu$~: inserting in equation (\ref{result-mu-g}) the trivial (hydrodynamic) approximation 
$$g_0(r)=1 \quad,\quad 0<r<\infty$$
would lead to a vanishing result for $\mu$ and to the superfluid kind of behaviour described by equations 
(\ref{correlator})-(\ref{area}).

We now turn our attention to the off-diagonal terms $\mu^{(ij)}_{\alpha\beta}$ which couple the internal modes
of vortices $i$ and $j$. We have~:
\begin{eqnarray}
\langle\mu^{(ij)}_{\alpha\beta}\rangle_0 & = & \frac{1}{Z_0}\int\prod_{k=1}^N d{\bf r}_{0k}
\big[-\partial_\alpha\partial_\beta V({\bf r}_{0i}-{\bf r}_{0j})\,\big]\,\mbox{e}^{-\beta H_0} \nonumber\\
& = & -\frac{1}{Z_0}\int d{\bf r}_{01}\,d{\bf r}_{02}
\;\partial_\alpha\partial_\beta V({\bf r}_{01}-{\bf r}_{02})\times
\nonumber\\
&\times&\int d{\bf r}_{03}\dots d{\bf r}_{0N} \,\mbox{e}^{-\beta H_0}
\nonumber
\end{eqnarray}
Comparing the {\em rhs} of the last equation with the {\em rhs} of equation (\ref{interm3}), we see that
\begin{eqnarray}
\langle\mu^{(ij)}_{\alpha\beta}\rangle = -\frac{1}{(N-1)}\,\langle\mu^{(i)}_{\alpha\beta}\rangle
\label{avg-offdiag}
\end{eqnarray}
which, in view of the fact that the mass $\mu=\langle\mu^{(i)}\rangle$ is finite
({\em i.e.} independent of $N$), 
implies that the off diagonal terms vanish in the thermodynamic limit $N\to\infty$.

We are now left with the following averaged Hamiltonian for the internal degrees of freedom of our flux line
liquid~:
\begin{eqnarray}
H_{eff} & = & \sum_{i=1}^N\int_0^L dz\,\,
\Big[\,\frac{1}{2}\,\tilde\varepsilon_1\,\Big(\frac{d{\bf u}_i}{dz}\Big)^2
+ \frac{1}{2}\,\mu\, {\bf u}_i^2(z) \,\Big] + \nonumber\\
& - & \frac{1}{(N-1)} \sum_{i=1}^N\sum_{j\neq i}\int_0^L dz\;\frac{1}{2}\mu\,{\bf u}_i(z)\cdot{\bf u}_j(z)
\label{H1tot}  
\end{eqnarray}
where we used the results (\ref{avg-diag})-(\ref{avg-offdiag}) for the averaged diagonal 
$\langle\mu^{(i)}_{\alpha\beta}\rangle$ and 
off-diagonal $\langle\mu^{(ij)}_{\alpha\beta}\rangle$ terms, in conjunction with equation (\ref{TaylorHam1}).
This last expression of $H_{eff}$ can be rewritten in Fourier space in the form
\begin{eqnarray}
H_{eff} & = & \sum_{i=1}^N \sum_{n\neq 0}\;\frac{1}{2}\,(G^{-1})_{ij}(q_n){\bf r}_i(q_n)\cdot{\bf r}_j(-q_n)
\label{Heff}
\end{eqnarray}
with the inverse propagator
\begin{eqnarray}
(G^{-1})_{ij}(q_n) = \big[\,L\tilde\varepsilon_1\,q_n^2 + L\mu + \frac{L\mu}{N-1}\,]\,\delta_{ij} - \frac{L\mu}{N-1}
\nonumber
\end{eqnarray}
From this last equation, the propagator $G_{ij}(q_n)$ defined through
\begin{eqnarray}
\langle u_i(q_n)u_j(q_m)\rangle = k_BTG(q_n)\delta_{n,-m}
\end{eqnarray}
can be easily obtained using an identity for inverting $N\times N$ matrices of the form
\begin{eqnarray}
(A^{-1})_{ij} = a\delta_{ij} + b
\end{eqnarray}
namely,
\begin{eqnarray}
A_{ij} = \frac{1}{a}\delta_{ij} - \frac{b}{a(a+bN)}
\end{eqnarray}
We thus obtain for $G_{ij}(q_n)$ the following result
\begin{equation}
G_{ij}(q_n) = \frac{1}{\Gamma(q_n)}\,\delta_{ij} + \frac{L\mu/(N-1)}{\Gamma(q_n)\big[\Gamma(q_n) - L\mu N/(N-1)\big]}
\label{G(N)}
\end{equation}
where we denote by $\Gamma(q_n)$ the quantity
\begin{eqnarray}
\Gamma(q_n) = \big[\,L\tilde\varepsilon_1\,q_n^2 + L\mu + \frac{L\mu}{N-1}\,]
\end{eqnarray}
In the thermodynamic limit $N\to\infty$, we see from equation (\ref{G(N)}) that $G_{ij}(q_n)$ reduces to the diagonal form
$G_{ij}(q_n)=G(q_n)\delta_{ij}$, with
\begin{eqnarray}
G(q_n) = \frac{1}{L\tilde\varepsilon_1\,q_n^2 + L\mu}
\end{eqnarray}
Hence, the total projected area of a given flux line in the liquid environment is given by
\begin{eqnarray}
\langle u^2(z)\rangle & = & d_\perp k_BT\sum_{n\neq 0} G(q_n) \nonumber\\
& = &\frac{2d_\perp k_BT}{L\tilde\varepsilon_1}\sum_{n=1}^\infty\frac{1}{q_n^2 + (\mu/\tilde\varepsilon_1)}
\label{massivewidth}
\end{eqnarray}
Using the result (\ref{resultmu}) in equation (\ref{massivewidth}), along with the fact that
\begin{eqnarray}
\sum_{n=1}^\infty\frac{1}{n^2 + a^2} & = & \frac{\pi}{2a}\,\mbox{coth}(\pi a) - \frac{1}{2a^2} \nonumber\\
&\simeq& \frac{\pi}{2a}
\quad\mbox{for}\quad a\to\infty\,\, ,
\end{eqnarray}
we obtain (we take $d_\perp=2$)
\begin{eqnarray}
\langle u^2\rangle \simeq \frac{k_BT}{2\varepsilon\varepsilon_0\,\sqrt{\pi\rho}}
\end{eqnarray}
The important thing to note about this result is that, unlike the free flux line, the mean square projected area
$\langle u^2\rangle$ is not proportional to and in fact does not depend at all on the sample thickness $L$. 
More significantly, if we take $\rho\approx 1/a^2$, where $a$ is the average intervortex distance in the liquid 
phase, the above equation gives us~:
\begin{eqnarray}
\langle u^2\rangle \approx \frac{k_BT\,a}{2\sqrt{\pi}\varepsilon\varepsilon_0}
\label{conf-u2}
\end{eqnarray}
which is exactly the result one obtains in the crystalline phase, equation (\ref{usqrd-crystal}), and is actually 
what we expect on purely physical grounds for a flux line trapped in the cage potential formed by its surrounding 
neighbors, assuming that the barriers for cutting and crossing are high enough.
That we are able to obtain such a result for $\langle u^2\rangle$ gives us confidence in our approach and in 
the various approximations that were made in the course of our calculation.

Having obtained $\langle u^2\rangle$, we are now in a position to assess the range of validity of our analysis.
Using the fact that \cite{Blatter-et-al} $\varepsilon_0(K/\AA)=1.964\times 10^8/[{\lambda}(\AA)]^2$, 
and the values $\varepsilon=1/7$, ${\lambda}(0)=1400\AA$ relevant to YBCO, we find 
\begin{eqnarray}
\frac{\langle u^2\rangle}{a^2} \simeq \frac{0.27\times 10^{-2}}{1-T/T_c}\,\, \Big(\frac{{\lambda}(0)}{a}\Big)
\label{resultu2}
\end{eqnarray}
where we approximated $\sqrt{\rho}\approx 1/a$, and where we used the mean field dependence of the London 
penetration depth on
temperature ${\lambda}(T)={\lambda}(0)/\sqrt{1-T/T_c}$. The width of the flux lines in a dilute liquid 
($a\lesssim{\lambda}$)
is hence very small compared to the average inter-vortex distance $a$. We need to go to much higher densities
(i.e. to a higher ratio ${\lambda}/a$) and to temperatures very close to $T_c$ in order to have
$\langle u^2\rangle\approx a^2$.
More quantitatively, in order for our perturbative analysis based on the taylor expansion (\ref{Taylorexpansion}) 
to be correct, the condition (\ref{condition}) has to be satisfied.
This condition can be rewritten in the following, averaged form~:
\begin{eqnarray}
\langle[{\bf u}_i(z)-{\bf u}_j(z)]^2\rangle < a^2 \quad,\quad \mbox{for all}\quad i\neq j
\end{eqnarray}
But, since $\langle{\bf u}_i(z)\cdot{\bf u}_j(z)\rangle=
\langle{\bf u}_i(z)\rangle\cdot\langle{\bf u}_j(z)\rangle=0$ for $i\neq j$ 
in our approach (we remind the reader 
that the off-diagonal coefficients $\mu^{(ij)}_{\alpha\beta}$ were found to vanish in the thermodynamic limit),
the condition above can be rewritten in the form
\begin{equation}
\langle {\bf u}_i^2\rangle < \,a^2/2 \label{condition2}
\end{equation}
which, by equation (\ref{resultu2}), is sure to be satisfied on a rather large region of the phase diagram $(H,T)$ of
the flux line system.

Let us now find the difference correlation function $\langle[{\bf u}_i(z)-{\bf u}_i(0)]^2\rangle$, which is given 
by~:
\begin{eqnarray}
\langle[{\bf u}(z)-{\bf u}(z')]^2\rangle = \frac{4d_\perp k_BT}{L\tilde\varepsilon_1}
\sum_{n=1}^\infty \,\frac{1-\cos q_n(z-z')}{q_n^2 + (\mu/\tilde\varepsilon_1)}
\nonumber
\end{eqnarray}
Transforming the sum into an integral, and using the result
$$\int_0^\infty dq\,\,\frac{1-\cos(qx)}{q^2+a^2} = \frac{\pi}{2a}\,\big[1-\mbox{e}^{-a|x|}\big] \qquad
,\qquad a>0$$
we obtain, after a few manipulations~:
\begin{eqnarray}
\langle[{\bf u}(z)-{\bf u}(z')]^2\rangle \simeq \frac{d_\perp k_BT a}{\sqrt{\pi}\varepsilon\varepsilon_0}
\big(1-\mbox{e}^{-\frac{\sqrt\pi}{\varepsilon a}\,|z-z'|}\big)
\label{conf-corr}
\end{eqnarray}
We thus find that the mean square displacement $\langle[{\bf u}(z)-{\bf u}(z')]^2\rangle$
goes to a finite limit as $|z-z'|\to\infty$. This was expected, since in our approach, the internal
modes are massive and have bounded fluctuations which do not grow with the sample thickness.

The last quantity we shall be interested in is the density-density correlation function
\begin{eqnarray}
S({\bf r}-{\bf r}';z-z') = \langle\hat\rho({\bf r},z)\hat\rho({\bf r}',z') \rangle 
\label{strucfac}
\end{eqnarray}
where $\hat\rho({\bf r},z)$ is the local density operator at height $z$
\begin{eqnarray}
\hat\rho({\bf r},z) = \sum_{i=1}^N\delta\big({\bf r}-{\bf r}_i(z)\big)
\end{eqnarray}
According to the theory of reference \cite{Nelson-Seung}, the partial Fourier transform 
$S({\bf q},z)$ should behave as 
\begin{eqnarray}
S({\bf q},z) \simeq S({\bf q},0)\,\mbox{e}^{-\varepsilon(q)|z|}
\end{eqnarray}
where $\varepsilon(q)$ is the Bogoliubov spectrum of the corresponding bosons. The above form of $S({\bf q},z)$
suggests that the densities at two different heights $z$ and $z'$ become more decorrelated as $|z-z'|$
grows. Here, due to the fact that the internal modes are massive, we expect the density to remain correlated on
the whole longitudinal length of the sample.
Indeed, in Appendix B we show that the structure factor $S({\bf q},z)$ within our model is given by~:
\begin{eqnarray}
S({\bf q},z) & = & \rho^2g_0({\bf q})\;
\mbox{e}^{-q^2 \frac{k_BT a}{2\eta\sqrt{\pi}\varepsilon\varepsilon_0}}
+ \nonumber\\
& + &\rho\exp\Big(-q^2\;\frac{k_BT a}{2\sqrt{\pi}\varepsilon\varepsilon_0}
\big(1-\mbox{e}^{-\frac{\sqrt\pi}{\varepsilon a}\,|z|}\big)
\,\Big)
\label{conf-struc-fac}
\end{eqnarray}
which shows that the structure factor consists of two pieces~: a CM piece, proportional to $\rho^2$, which
corresponds to a liquid of more or less straight flux lines and includes, through $g_0({\bf q})$, nontrivial 
correlations between the CM positions of flux
lines, and a second piece linear in the density $\rho$, which describes the internal 
fluctuations of the individual lines and which incorporates the effect of 
interactions between vortices through the 
confining mass $\mu$.
Like the structure factor of the boson analogy, our $S({\bf q},z)$
does decrease as a function of $|z|$, with the important difference, however, that
$S({\bf q},z)$ here goes to a finite limit (much like in a crystal) as $|z|$ grows very large,
in contrast to the result of ref.\cite{Nelson-Seung} where $S({\bf q},z)\to 0$ as $|z|\to\infty$,
indicating that flux lines remain correlated over rather long length scales in a disentangled flux line liquid.

The most important implication of the theory developped in this section is related to the issue of entanglement
of flux lines. {\em Assuming} that the typical energy scale for the interactions between flux lines is much
higher than $k_BT$ and neglecting thermally nucleated vortex loops, 
we have derived the statistical mechanics of 
a dilute line liquid using a simple Taylor expansion of the Hamiltonian in terms of the internal modes of the 
flux lines. We have calculated the mean square projected area of a given flux line $\langle u^2\rangle$
and we have found that it is of the same order of magnitude as what is usually 
obtained in a flux line lattice. The physical picture of a flux line in a liquid environment that emerges from our
analysis, under the above assumptions, 
is that of a roughly straight object whose internal modes are contained in a
tube of finite radius $\langle u^2\rangle\le a$, no matter how thick the sample might be.
At this stage, we can already assert that we have achieved the goal stated in the Introduction, and that we now 
have at our disposal a detailed and yet simple theory which is able to describe a 
completely {\em disentangled} flux 
line liquid. As we mentioned earlier, the assumption $E_c\gg k_BT$ is, however, too strong for real 
flux liquids on the melting line $H_m(T)$ (as can be easily checked using material parameters relevant to most 
HTSC), and our theory needs to be adjusted to take this fact into account. 
In the following section, we 
discuss and compare our findings with previous work, and we try to construct a consistent picture of what we 
expect the physics of flux line liquids to be based on our approach.

\section{Discussion and Conclusions}
\label{Conclusions}

We shall start this section by briefly reviewing and discussing 
the results of numerical simulations of flux line liquids. A number of these simulations 
have used the mapping of the Ginzburg-Landau free energy of a superconductor onto the Hamiltonian of the uniformly 
frustrated three dimensional (3D) XY model~:
\begin{eqnarray}
H_{XY} = -\sum_{\langle{i,j}\rangle} J_{i,j}\,\cos(\theta_i-\theta_j-A_{ij})
\label{3DXY}
\end{eqnarray}
where, in a discrete lattice, $\theta_i$ is the phase of the complex Ginzburg-Landau order parameter
$\psi({\bf r},z)=|\psi({\bf r},z)|\,\mbox{e}^{i\theta({\bf r},z)}$ at site $i$, 
$A_{ij}=(2\pi/\phi_0)\int_{i}^{j}{\bf A}\cdot{d\bf l}$ is the line integral of the magnetic vector potential along 
a path linking site $i$ to site $j$, and the sum is over nearest-neighbor sites.
The Hamiltonian (\ref{3DXY}) results from making the London approximation 
$\psi_0=|\psi_0({\bf x},z)|=\mbox{Cst.}$ in the discretized 
Ginzburg-Landau free energy of an anisotropic superconductor in the regime ${\lambda}/a\to\infty$, so that the 
magnetic induction ${\bf B}=B\hat{\bf z}$ inside the superconductor can be taken as a constant.
For a layered superconductor with the principal axis ${\bf c}$ along the $\hat{\bf z}$ axis,
the couplings $J_{i,j}$ are given by $J_{i,j}=J_\perp$ when sites $i$ and $j$ are located on the same 
superconducting plane, with $J_\perp=\phi_0^2s/(16\pi^3{\lambda}^2)$, and 
$J_{i,j}=J_z=\varepsilon\phi_0^2\xi^2/(16\pi^3s{\lambda}^2)$ 
when sites $i$ and $j$ are located at $\pm s\hat{\bf z}$ from each other, where $s$ is the spacing between 
superconducting layers.

Within the model (\ref{3DXY}), evidence for an entangled state is deduced from 
various dynamical quantities, such as the helicity modulus $\Upsilon({\bf q})$ which gives the linear response
of the supercurrent ${\bf j}$ to a transverse perturbation in the vector potential ${\bf A}^{ext}$ of the 
externally applied magnetic field~:
\begin{eqnarray}
j_\mu(q_\nu) = -\Upsilon_\mu(q_\nu)\;\delta{A}_{\mu}^{ext}(q_\nu) \quad,\quad \mu\neq\nu
\end{eqnarray}
The helicity modulus measures the response of the system to an imposed phase 
twist\cite{Fisher-et-al-73,Ohta-79,Miyashita-84}, 
and is therefore related to the correlation function
(here ${\bf R}=({\bf r},z)$)
\begin{eqnarray}
\Psi({\bf R})=\Big\langle\psi({\bf R})\psi^*({\bf 0})
\exp\big(-\frac{2i\pi}{\phi_0}\int_{\bf 0}^{\bf R}{\bf A}\cdot{d\bf l}\big)
\Big\rangle
\label{Psi-phase}
\end{eqnarray}
which measures the coherence of the phase degrees of freedom in the system.
Also measured in numerical simulations of model (\ref{3DXY}) is the
structure factor $S({\bf r},z)=\langle\hat\rho({\bf r},z)\hat\rho({\bf 0},0)\rangle$, 
where the density operator $\hat\rho({\bf r},z)=\sum_{i=1}^N\delta\big({\bf r}-{\bf r}_i(z)\big)$ is given in 
terms of the phase variables $\theta({\bf r},z)$ by
$$\hat\rho({\bf x},z)=\frac{1}{2\pi}\;\big[\nabla\times\nabla\theta({\bf x},z)\,\big]\cdot\hat{\bf z}$$
Information about entanglement and longitudinal correlations is then deduced from the 
behaviour of the structure factor 
$S({\bf 0},z)=\langle\hat\rho({\bf r},z)\hat\rho({\bf r},0)\rangle$ at zero transverse separations.

Another model which has been used in numerical studies of flux liquids is the so-called lattice London model, 
which is defined by the 
Hamiltonian\cite{Carneiro-et-al-prb-93,Chen-Teitel-prl-95,Nguyen-Sudbo-Hetzel-prl-96}
\begin{eqnarray}
H = 2\pi^2J_\perp\sum_{i,j}\sum_{\mu=x,y,z} 
G_\mu({\bf R}_i-{\bf R}_j)n_\mu({\bf R}_i)n_\mu({\bf R}_j)
\label{Lattice-London}
\end{eqnarray}
where $n_\mu({\bf R}_i)$ is the integer vorticity through plaquette $\mu$ at site 
${\bf R}_i=({\bf r}_i,z_i)$ of a cubic mesh of 
points, and $J_\perp$ is the same coupling as the one defined above in the context of the uniformly frustrated 3D 
XY model. The lattice London interactions $G_\mu$ have Fourier transforms~:
\begin{eqnarray}
G_{x,y}({\bf q}) & = & \frac{\varepsilon^2}{Q_x^2 + Q_y^2 + \varepsilon^2\,(Q_z^2 + s^2/{\lambda}^2)}
\nonumber\\
G_z({\bf q}) & = & \frac{(Q^2+\varepsilon^2s^2/{\lambda}^2)/(Q^2+s^2/{\lambda}^2)}
{Q_x^2 + Q_y^2 + \varepsilon^2\,(Q_z^2 + s^2/{\lambda}^2)}
\end{eqnarray}
where $Q_\mu=2\sin(q_\mu s/2)$ and $Q^2=\sum_\mu Q_\mu^2$. Unlike the 3D XY model, which has Coulomb like 
interactions between vortex elements\cite{Li-Teitel-prl-91} and which is only valid in the regime ${\lambda}\gg a$ 
of dense 
systems, the lattice London model takes screening into account and can therefore be used for dense as well as for 
dilute flux liquids.

The last kind of simulation which has been performed is the 
Monte Carlo simulation of Nordborg and Blatter\cite{Nordborg-Blatter-prl-97,Nordborg-Blatter-prb-98}
who use the boson mapping\cite{Nelson-Seung} to simulate a system of interacting flux lines. 
In this approach, the authors adapt an algorithm which has been used in the past\cite{Ceperley} to study 
superfluidity in quantum-mechanical Bose systems, involving large scale permutations of vortex trajectories 
subject to the ``periodic'' boundary conditions ${\bf r}_i(0)={\bf r}_j(L)$, {\em i.e.} every line ends either on 
itself or on another line. Below we comment on this Monte Carlo method and the associated algorithm, 
but, before doing so, we want to consider the two other simulation methods, models (\ref{3DXY}) and 
(\ref{Lattice-London}).

We first observe that 
a direct comparison between our results and simulations of the uniformly frustrated 3D XY model (\ref{3DXY}) is 
made difficult by the fact that, while this model is relevant to the case ${\lambda}\gg a$,
we here consider the opposite limit $a\lesssim{\lambda}$ of a dilute flux line liquid.
In addition, and as we already mentioned, most of the simulations above were concerned 
with the measurement of the 
helicity modulus, which is essentially a measure of {\em phase} correlations in the system. It should be noted, 
however, that phase correlations can be very different from density correlations.
To illustrate this statement, we refer the reader 
to ref.\cite{Glazman-Koshelev}, where it was found within a simple elastic approach
that the order parameter for phase correlations 
(\ref{Psi-phase}) decays exponentially in the direction of the superconducting planes 
in a three dimensional flux line {\em lattice}\cite{Remark3}, 
although the density order parameter 
$S_{\bf K}({\bf r},z)=\langle\hat\rho_{\bf K}({\bf r},z)\hat\rho^*_{\bf K}({\bf 0},0)\rangle$ is known to
be finite in such a lattice. 
This example shows clearly that there is no ``one to one'' correspondence between phase 
and density correlations, and that the vanishing of the helicity 
modulus in a flux line liquid for example does not 
necessarily mean exponential decay of density correlations along the $\hat{\bf z}$ axis or 
correlations between flux line positions of the type of equations (\ref{correlator})-(\ref{area}).
This is even more so if we allow for the formation of vortex loops. Thermally excited vortex loop 
excitations were found \cite{Nguyen-Sudbo-Hetzel-prl-96} to destroy phase coherence and to lead to a vanishing of 
the helicity modulus even in an otherwise perfectly ordered flux line lattice\cite{Frey-et-al-prb-94}.

Finally, it should be realized that the simulations of models (\ref{3DXY}) and (\ref{Lattice-London}) are 
highly simplified representations of considerably more complex physics. For example, 
simulations are usually carried out at a given density of flux quanta 
$f=(B\xi^2/\phi_0)=(B/2\pi H_{c2})$ 
(we remind the reader that the upper critical field for ${\bf H}||{\bf c}$ is given by $H_{c2}=\phi_0/2\pi\xi^2$), 
with values of $f$ ranging from $1/12$ to $1/30$ most frequently used. Values of $f$ in this range 
correspond to values of the induction $B$ in the range $\sim\frac{1}{2}H_{c2}-\frac{1}{5}H_{c2}$. 
Apart from the fact that the use of models (\ref{3DXY}) and (\ref{Lattice-London}), which are based on the 
assumption of a constant amplitude of the order parameter, becomes questionable in this range of fields, we also 
note that at such high fields vortices in adjacent superconducting 
planes in the liquid phase might very well be already decoupled\cite{Glazman-Koshelev} 
(note also that the temperature dependence of
the coherence length $\xi$ in $f=(B\xi^2/\phi_0)$ and of the London penetration depth ${\lambda}$ in $J_\perp$ and 
$J_z$ is generally left out as $T$ is varied).
Moreover, in most of these simulations, 
which are carried out on a cubic mesh, the mesh constant in the $\hat{\bf z}$ direction is associated with
the distance $s$ between superconducting planes in a layered material, and 
no mention is made whatsoever of the coherence length $\xi_c$ along the ${\bf c}$ axis. In particular, the 
question regarding whether the same results, {\em e.g.} for density correlations in the liquid 
phase, would follow in the regimes $\xi_c(T)<s$ and $\xi_c(T)>s$, has largely remained untouched. 
The fact that most of these simulations consider {\em implicitely} the 
quasi-two-dimensional case\cite{Klemm-et-al-75} 
$\xi_c(T)<s$ does not allow us to draw any conclusions on the interesting regime $\xi_c(T)\gg s$ where 
the average tilt angle of flux lines\cite{Nelson-Vinokur} 
$\langle(d{\bf r}_i/dz)^2\rangle\sim \big(k_BT/\varepsilon^2\varepsilon_0\xi_c(T)\big)$ 
can be very small and where
we expect superconducting coherence, whether it be for the phase of the superconducting order parameter or for 
the density, to survive on much longer length scales.

Several remarks are due regarding the Monte Carlo simulation of Nordborg and 
Blatter\cite{Nordborg-Blatter-prl-97,Nordborg-Blatter-prb-98} who use the boson mapping\cite{Nelson-Seung} to 
simulate a system of interacting flux lines.
First, the use of the ``periodic'' boundary conditions ${\bf r}_i(L)={\bf r}_j(0)$ 
imposes additional constraints on the system which are {\em not} 
present in a real flux line liquid. Actually, in order to ``capture the effects of Bose statistics'', the flux 
lines in this simulation were made to switch their endpoints {\em by hand}, 
and this might very well
introduce an {\em artificial} entanglement in the system. More specifically, the switching of endpoints 
is achieved by cutting out sufficiently long segments of a number of lines and trying different ways of connecting 
the loose ends while satisfying the periodic boundary conditions\cite{Remark2}.
While this procedure and the corresponding algorithm are appropriate for the imaginary time paths of
quantum mechanical bosons, we find it more realistic that the flux line system should be 
allowed to equilibrate with free (as opposed to periodic) boundary conditions and, more importantly, without 
explicit switching of endpoints. 
Within the context of quantum mechanical bosons, superfluidity is brought about in numerical simulations by
{\em precisely} these kind of manipulations. In the context of vortices,
we expect these manipulations to lead to an overestimation of the effect of entanglement in flux liquids,
and the ``superfluid'' behaviour found in reference \cite{Nordborg-Blatter-prb-98} 
might therefore be a direct consequence of 
the specific bosonic algorithm used in this simulation.

In view of all the above remarks, and the results of section \ref{Theory-disentangled}, 
it seems to us that there is still room, 
at least in the dilute limit considered in this paper and  neglecting vortex loops,
for a new phase of the flux line liquid which has not been 
considered in the past, and which we might describe as {\em weakly entangled}.
By ``weakly entangled'' flux liquid, we mean a liquid phase 
in which $\langle u^2\rangle$ can be very large, but does not actually diverge with the sample thickness $L$. 
For this to happen, a small value of the confining mass $\mu$, much smaller than the one found in equation 
(\ref{resultmu}), is needed. 
In the following paragraph, we try to understand how such small values of $\mu$ can emerge from our model, and how 
a weakly entangled phase can be reconciled with what is observed in numerical simulations.

The key quantity in the derivation of the confining mass of section \ref{Theory-disentangled} is the 
pair distribution function $g_0({\bf r})$ of the CM mode. 
In equation (\ref{pair-distr-func}), we used an approximation for 
$g_0(r)$ which is relevant to a situation where the average distance between the centers of mass of the flux lines 
is $a=1/\sqrt{\rho}$, and where the CM positions are strongly anti-correlated for $r\leq a$, 
{\em i.e.} if the CM of the $i$th flux line is at location ${\bf r}_{0i}$, then the centers of mass 
of neighboring vortices have a very small probability to be within a distance $a$ from ${\bf r}_{0i}$. 
While this is a perfectly legitimate way of thinking for actual flux line elements and for the regime
$E_c(T)\gg k_BT$ considered in section \ref{Theory-disentangled},
since the CM is only a mathematical construct on one hand, and since the condition $E_c(T)\gg k_BT$ 
is usually not satisfied near melting on another
(making situa-
\begin{minipage}[t]{3.2in}
\vspace{0.1in}
\epsfxsize=2.2in
\hfill \epsfbox{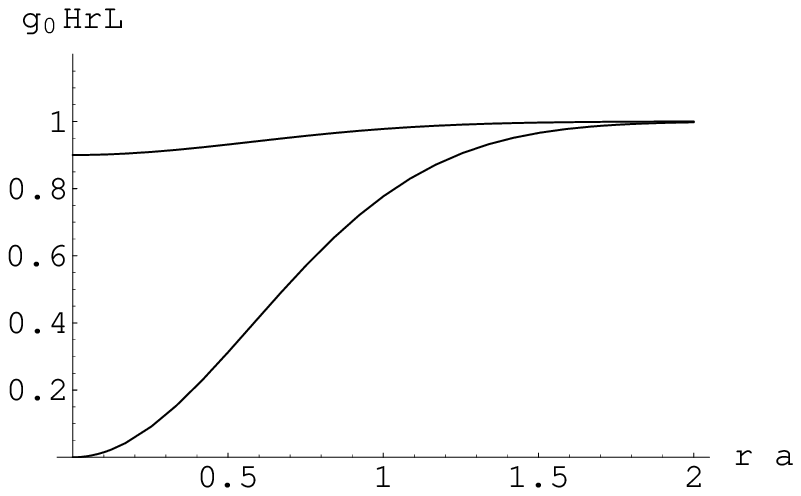} \hfill
\begin{small}   

Figure 2. Pair distribution function $g_0(r)$ of equation (\ref{mod-g0}) for $\alpha = 3/2$ and $\eta=0.1$
(upper curve). For comparison, we also show the pair correlation function of equation
(\ref{pair-distr-func}), (lower curve).
\end{small}
\end{minipage}

\medskip

\noindent
tions where CM positions of different flux lines are very close to each other rather frequent), 
the CM distribution function does {\em not} have to be very small ($g_0(r)\ll 1$) for $r\leq a$. 
However, because of the repulsion between flux lines, we cannot {\em completely} neglect 
correlations between the center of mass positions and simply take $g_0(r)=1$ for $0<r<\infty$, in which case we 
would recover a fully entangled line liquid with the correlations (\ref{correlator})-(\ref{area}).
Since there is obviously no easy way to derive an analytic expression for the pair distribution function $g_0(r)$, 
we here are bound to speculate about its possible shape using our physical intuition and infer the resulting 
physics. One possible shape for $g_0(r)$ which allows for close encounters between the centers of mass of 
different flux lines is the one shown in figure 2, and corresponds to the approximation
\begin{eqnarray}
g_0(r) \simeq 1 -\eta\,\exp\big(-\alpha r^2/a^2\big)
\label{mod-g0}
\end{eqnarray}
where $\eta$ is a numerical constant such that 
$0< \eta < 1$ (in figure 2, we use $\eta = 0.1$). 
Such a form of $g_0(r)$ would yield a confining mass which is $\eta$ times smaller than the 
one found previously, equation (\ref{resultmu}), and would therefore lead to values of the average projected area
$\langle u^2\rangle$ 
\begin{eqnarray}
\langle u^2\rangle \approx \frac{k_BT\,a}{2\eta\sqrt{\pi}\varepsilon\varepsilon_0}
\label{gen-conf-u2}
\end{eqnarray}
which are much larger than the square of the average distance $a$ between flux lines. 
On the other hand, equation (\ref{mod-g0}) would lead to a relative displacement
\begin{eqnarray}
\langle[{\bf u}(z)-{\bf u}(z')]^2\rangle \simeq \frac{d_\perp k_BT a}{\eta\sqrt{\pi}\varepsilon\varepsilon_0}
\big(1-\mbox{e}^{-\frac{\eta\sqrt\pi}{\varepsilon a}\,|z-z'|}\big)
\label{gen-conf-corr}
\end{eqnarray}
For $\eta\ll 1$ there will be a substantial range of separations 
\begin{eqnarray}
|z-z'|< \ell_z=\frac{\varepsilon a}{\eta\sqrt\pi}
\label{ellz}
\end{eqnarray}
where the {\em rhs\ } of equation (\ref{gen-conf-corr}) can very well be approximated by
\begin{eqnarray}
\langle[{\bf u}(z)-{\bf u}(z')]^2\rangle \simeq 
\frac{d_\perp k_BT}{\varepsilon^2\varepsilon_0}\;|z-z'|
\end{eqnarray}
(note that factors of $\eta$ have cancelled each other in this last equation)
and for which density correlations $\langle\hat\rho({\bf q},z)\hat\rho(-{\bf q},0)\rangle$
will decrease exponentially as a function of $|z|$, leading to the structure factor
\begin{eqnarray}
S({\bf q},z) \simeq \rho\,\mbox{e}^{-q^2\;\frac{d_\perp k_BT}{\varepsilon^2\varepsilon_0}\;|z|}
\label{rho-rho-f}
\end{eqnarray}
Note, in particular that for moderately anisotropic superconductors, and if $\eta\ll 1$, the length $\ell_z$ can
be quite large, maybe even larger than the relatively small thicknesses used in numerical simulations,
giving the {\em semblance} that the entangled state is characterized by the correlations 
(\ref{correlator})-(\ref{area}) with an entanglement correlation length
\begin{eqnarray}
\xi_v\simeq \frac{\varepsilon^2\varepsilon_0a^2}{d_\perp\,k_BT}\qquad .
\nonumber
\end{eqnarray}
On the other hand, for values of $\eta$ close to unity ($\eta\lesssim 1$), we recover a flux line liquid where 
correlations extend over much longer distances, which corresponds to the situation analyzed by Righi et 
{\em al.}\cite{Righi-et-al}. We thus see that our model can describe, depending on what $g_0(r)$ in reality is, 
very different situations corresponding to nearly disentangled ($\eta\lesssim 1$), 
weakly entangled ($\eta\ll 1$) or fully entangled ($\eta=0$, $g_0(r)=1$) flux line liquids.

To summarize, in this paper, using a perturbative expansion of the Hamiltonian of interacting flux lines in a 
vortex liquid similar to the expansion of the action of a quantum particle around a classical 
path\cite{Feynman-Hibbs}, we have constructed a mean field theory of the flux line liquid which, in the
author's opinion, has a better handling of the internal fluctuations of flux lines,
and which, through the use of a nontrivial pair distribution function $g_0(r)$ for the positions of the centers 
of mass of flux lines, goes beyond Gaussian hydrodynamics,
making contact with the standard liquid state theory of classical fluids.
Within our approach, we find that a weakly entangled phase might exist, where the 
average width of flux lines $\langle u^2\rangle^{\frac{1}{2}}$ can be
much larger than the average intervortex distance $a$, but does 
not diverge with the sample thickness $L$. By varying the shape of $g_0(r)$, 
we are able to describe situations with 
very short and rather large entanglement correlation lengths.
Since the slightest deviation of the CM pair distribution function from unity at small distances ($r\leq a$)
will necessarily lead to a finite (nonzero) confining mass 
$\mu$, a {\em careful} measurement of $g_0(r)$, for example in numerical simulations, 
would provide an unequivocal way of verifying the predictions of this paper, and to ascertain
whether a fully entangled liquid phase with correlations 
(\ref{correlator})-(\ref{area}) or the weakly entangled phase proposed here is the 
correct ground state of the flux line liquid.

\acknowledgments

The author wishes to thank Professor Leo Radzihovsky for useful remarks and for criticism of the 
first version of this manuscript. This work was supported by the NSF through grant DMR--9625111.

\section{Appendix A~:~Details of the calculation of the mass $\mu$ for the internal modes of vortices 
in a flux line liquid }

In this appendix, we show details of the calculation of the integral on the {\em rhs} of equation
(\ref{equation-mu}) which
gives the value of the effective mass $\mu$. Using the expression (\ref{interaction-k}) of the interaction
potential in
Fourier space $V(\bf k)$, and performing the integration over the polar angle, we obtain
\begin{eqnarray}
\mu & = & \frac{2\pi}{d_\perp\alpha}(\rho a^2)\int_0^\infty kdk\;\frac{\varepsilon_0k^2}{k^2+{\lambda}^{-2}}
\,\mbox{e}^{-k^2a^2/2\alpha}
\end{eqnarray}
Using the change of variables $x = {\lambda}^2k^2$, this last expression can be cast in the form
\begin{eqnarray}
\mu = \frac{\pi a^2}{d_\perp\alpha{\lambda}^2}\,(\rho\varepsilon_0)\int_0^\infty dx\;
\frac{x}{1+x}\,\mbox{e}^{-\kappa x}
\end{eqnarray}
with $\kappa = a^2/2\alpha{\lambda}^2$.
Now, using the result \cite{Prudnikov}
\begin{eqnarray}
\int_0^\infty dx\;\frac{x}{1+x}\,\mbox{e}^{-\kappa x} =
\frac{1}{\kappa} - \mbox{e}^\kappa\mbox{E}_1(\kappa)
\end{eqnarray}
where $\mbox{E}_1(x)=\int_x^\infty dt\,\mbox{e}^{-t}/t$ is the exponential integral,
and using the Taylor expansion for small arguments\cite{Abramowitz}
\begin{eqnarray}
\mbox{E}_1(x) = -\gamma -\ln x -\sum_{n=1}^\infty\frac{(-1)^nx^n}{n(n!)}
\end{eqnarray}
(here $\gamma=0.577\dots$ is Euler's constant), we obtain
\begin{eqnarray}
\mu = \frac{\pi a^2}{d_\perp\alpha{\lambda}^2}\,(\rho\varepsilon_0)\cdot
\frac{1}{\kappa}\big(1 + \kappa\mbox{e}^\kappa(\gamma + \ln\kappa) + o(\kappa^2)\big)
\end{eqnarray}
Using the fact that $\kappa = a^2/2\alpha{\lambda}^2 < 1$ in the regime 
$a<{\lambda}$ (assuming that $\alpha>1$), we finally obtain
\begin{eqnarray}
\mu &\simeq& \frac{\pi a^2}{d_\perp\alpha{\lambda}^2}\,(\rho\varepsilon_0)\cdot\frac{2\alpha{\lambda}^2}{a^2}
\nonumber\\
& = & \frac{2\pi}{d_\perp}\rho\varepsilon_0
\end{eqnarray}
which is the result quoted in the text, equation (\ref{resultmu}).

\section{Appendix B~:~Details of the calculation of the structure factor of a flux line liquid}

In this appendix, we show some details of the calculation of the structure factor
\begin{eqnarray}
S({\bf r},z;{\bf r}',z') =\langle\hat\rho({\bf r},z)\hat\rho({\bf r}',z')\rangle
\end{eqnarray}
both for a free flux line liquid of noninteracting flux lines, and for the weakly entangled flux line liquid 
described in the text. In general,
we expect flux liquids at equilibrium to be translationally invariant, so that the structure factor
will depend only on the relative coordinates $({\bf r}-{\bf r}')$ and $(z-z')$, {\em i.e.} 
$S({\bf r},z;{\bf r}',z')=S({\bf r}-{\bf r}',z-z')$. As a consequence, we have for the Fourier transform of the 
density-density correlation function~:
\begin{equation}
\langle\hat\rho({\bf q},q_z)\hat\rho({\bf q}',q_z')\rangle =
(2\pi)^d\delta({\bf q}+{\bf q}')\delta(q_z+q_z')\;S({\bf q},q_z)
\label{S(q)}
\end{equation}
In the following, we shall be concerned with the quantity
\begin{eqnarray}
S({\bf q},q_z) = \frac{1}{LL_\perp^{d_\perp}}\,\langle\hat\rho({\bf q},q_z)\hat\rho(-{\bf q},-q_z)\rangle
\end{eqnarray}
where we used equation (\ref{S(q)}) above and the fact that 
$(2\pi)^d\delta({\bf q}={\bf 0})\delta(q_z=0)\equiv LL_\perp^{d_\perp}$ in the limit 
$L,L_\perp\to\infty$. We have~:
\end{multicols}
\begin{eqnarray}
\langle\hat\rho({\bf q},q_z)\hat\rho(-{\bf q},-q_z)\rangle & = &
\int d{\bf r}dz\int d{\bf r}'dz'\;\langle\hat\rho({\bf r},z)\hat\rho({\bf r}',z')\rangle\;
\mbox{e}^{-i{\bf q}\cdot({\bf r}-{\bf r}')}\mbox{e}^{-iq_z(z-z')}
\nonumber\\
& = & \sum_{i=1}^N\sum_{j=1}^N\int d{\bf r}dz\int d{\bf r}'dz'\;
\langle\delta({\bf r}-{\bf r}_i(z))\delta({\bf r}'-{\bf r}_j(z'))\rangle\;
\mbox{e}^{-i{\bf q}\cdot({\bf r}-{\bf r}')}\mbox{e}^{-iq_z(z-z')}
\nonumber\\
&=&\sum_{i=1}^N\sum_{j\neq i}\int dz\int dz'\;
\langle\mbox{e}^{-i{\bf q}\cdot({\bf r}_{0i}-{\bf r}_{0j})}\rangle_0
\langle\mbox{e}^{-i{\bf q}\cdot({\bf u}_{i}(z)-{\bf u}_{j}(z'))}\rangle_1\;\mbox{e}^{-iq_z(z-z')} +
\nonumber\\
&+& \sum_{i=1}^N\int dz\int dz'
\langle\mbox{e}^{-i{\bf q}\cdot({\bf u}_{i}(z)-{\bf u}_{j}(z'))}\rangle_1\;\mbox{e}^{-iq_z(z-z')}
\nonumber\\
&=&\sum_{i=1}^N\sum_{j\neq i}\int dz\int dz'\;
\langle\mbox{e}^{-i{\bf q}\cdot({\bf r}_{0i}-{\bf r}_{0j})}\rangle_0                  
\;\mbox{e}^{-\frac{1}{2}q_\alpha q_\beta
\langle[u_{i,\alpha}(z)-u_{j,\alpha}(z')][u_{i,\beta}(z)-u_{j,\beta}(z')]\rangle_1}
\;\mbox{e}^{-iq_z(z-z')} +
\nonumber\\
&+& \sum_{i=1}^N\int dz\int dz'
\mbox{e}^{-\frac{1}{2}q_\alpha q_\beta
\langle[u_{i,\alpha}(z)-u_{i,\alpha}(z')][u_{i,\beta}(z)-u_{i,\beta}(z')]\rangle_1}
\;\mbox{e}^{-iq_z(z-z')}
\label{interm-1}
\end{eqnarray}
where $\langle\cdots\rangle_0$ and $\langle\cdots\rangle_1$ denote averages over the CM and internal modes 
with statistical weights $\exp(-\beta H_0)$ and $\exp(-\beta H_{eff})$ respectively 
($H_0$ and $H_{eff}$ are the Hamiltonians given in equations (\ref{H0}) and (\ref{H1tot}) of the text).
Both for the liquid of noninteracting lines, and for the weakly entangled liquid that we consider here, the 
internal degrees of freedom ${\bf u}_i(z)$ and ${\bf u}_j(z)$ belonging to two different lines $i\neq j$ are 
decoupled. We therefore can write, for $i\neq j$~:
\begin{eqnarray}
\langle[u_{i,\alpha}(z)-u_{j,\alpha}(z')][u_{i,\beta}(z)-u_{j,\beta}(z')]\rangle_1 
&=& \langle u_{i,\alpha}(z)u_{i,\beta}(z) + u_{j,\alpha}(z')u_{j,\beta}(z')\rangle_1
\nonumber\\
& = & \frac{2\delta_{\alpha\beta}}{d_\perp}\langle u^2\rangle
\end{eqnarray}
where, in going from the first to the second line, we used the fact that
$\langle u_{i,\alpha}u_{i,\beta}\rangle_1=\delta_{\alpha,\beta}\langle u_{i,\alpha}^2\rangle$.
Equation (\ref{interm-1}) becomes
\begin{eqnarray}
\langle\hat\rho({\bf q},q_z)\hat\rho(-{\bf q},-q_z)\rangle & = &
\sum_{i=1}^N\sum_{j\neq i}\int dz\int dz'\;
\langle\mbox{e}^{-i{\bf q}\cdot({\bf r}_{0i}-{\bf r}_{0j})}\rangle_0
\mbox{e}^{-\frac{1}{d_\perp}q^2\langle u^2\rangle}
\;\mbox{e}^{-iq_z(z-z')} +
\nonumber\\
&+& \sum_{i=1}^N\int dz\int dz'
\mbox{e}^{-\frac{1}{2}q_\alpha q_\beta
\langle[u_{i,\alpha}(z)-u_{i,\alpha}(z')][u_{i,\beta}(z)-u_{i,\beta}(z')]\rangle_1}
\;\mbox{e}^{-iq_z(z-z')}
\label{interm-2}
\end{eqnarray}
Using the fact that 
$\int dz\!\int dz' \;\mbox{e}^{-iq_z(z-z')} = L^2\delta_{q_z,0}$, 
and noticing that
\begin{eqnarray}
\sum_{i=1}^N\sum_{j\neq i}\langle \mbox{e}^{-i{\bf q}\cdot{\bf r}_{0i}}
\mbox{e}^{-i{\bf q}'\cdot{\bf r}_{0j}}\rangle_0 =
(2\pi)^{d_\perp}\delta({\bf q}+{\bf q}')\;\rho^2\,g_0({\bf q})
\end{eqnarray}
which gives us here (with ${\bf q}'=-{\bf q}$)
\begin{eqnarray}
\sum_{i=1}^N\sum_{j\neq i}\langle \mbox{e}^{-i{\bf q}\cdot{\bf r}_{0i}}
\mbox{e}^{i{\bf q}\cdot{\bf r}_{0j}}\rangle_0 = 
(2\pi)^{d_\perp}\delta({\bf q}=0)\;\rho^2\,g_0({\bf q})
= L_\perp^{d_\perp}\rho^2 g_0({\bf q})
\end{eqnarray}
we finally obtain~:
\begin{eqnarray}
\langle\hat\rho({\bf q},q_z)\hat\rho(-{\bf q},-q_z)\rangle & = &
L^2\delta_{q_z,0}\;L_\perp^{d_\perp}\,\rho^2g_0({\bf q})\;
\mbox{e}^{-\frac{1}{d_\perp}q^2\langle u^2\rangle} 
+ \sum_{i=1}^N\int dz\int dz'
\mbox{e}^{-\frac{1}{2d_\perp}q^2
\langle[{\bf u}_{i}(z)-{\bf u}_{i}(z')]^2\rangle_1}
\;\mbox{e}^{-iq_z(z-z')}
\label{interm-3}
\end{eqnarray}
where we used the fact that 
$\langle[u_{i,\alpha}(z)-u_{i,\alpha}(z')][u_{i,\beta}(z)-u_{i,\beta}(z')]\rangle_1 = 
\frac{\delta_{\alpha,\beta}}{d_\perp}\;\langle[{\bf u}_i(z)-{\bf u}_i(z')]^2\rangle_1$.
From this point on, things will differ depending on whether we consider a liquid of free flux lines or a weakly 
entangled flux liquid. We shall treat both cases separately, starting with a liquid of free flux lines.

\subsection{Liquid of free flux lines}

In a liquid of free flux lines (or in a fully entangled ``superfluid'' state, for that matter), the average width 
of flux lines $\langle u^2\rangle^{\frac{1}{2}}$ 
diverges with the sample thickness $L$. We can therefore drop the first term on 
the {\em rhs} of equation (\ref{interm-3}), which contains the factor 
$\mbox{e}^{-\frac{1}{d_\perp}q^2\langle u^2\rangle}$. If we in addition use the fact that
\begin{eqnarray}
\langle[{\bf u}_{i}(z)-{\bf u}_{i}(z')]^2\rangle_1 = \frac{d_\perp k_BT}{\tilde\varepsilon_1}\;|z-z'|
\end{eqnarray}
we obtain~:
\begin{eqnarray}
\langle\hat\rho({\bf q},q_z)\hat\rho(-{\bf q},-q_z)\rangle & = &
\sum_{i=1}^N\int dz\int dz'
\mbox{e}^{-\frac{1}{2}q^2
\frac{k_BT}{\tilde\varepsilon_1}\;|z-z'|}
\;\mbox{e}^{-iq_z(z-z')}
\nonumber\\
& = & NL\int_{-\infty}^{\infty} dZ\;
\mbox{e}^{-\frac{1}{2}q^2
\frac{k_BT}{\tilde\varepsilon_1}\;|Z|}          
\;\mbox{e}^{-iq_z Z}
\label{interm-2-1}
\end{eqnarray}
Performing the $Z$ integration, and using the fact that the density $\rho=N/L_\perp^{d_\perp}$, we finally obtain
\begin{eqnarray}
\langle\hat\rho({\bf q},q_z)\hat\rho(-{\bf q},-q_z)\rangle = LL_\perp^{d_\perp}\;
\frac{\rho k_BTq^2/\tilde\varepsilon_1}{q_z^2 + \big(k_BTq^2/2\tilde\varepsilon_1\big)^2}
\end{eqnarray}
From the structure factor $S({\bf q},q_z)=\langle\hat\rho({\bf q},q_z)\hat\rho(-{\bf q},-q_z)\rangle/
LL_\perp^{d_\perp}$, we can deduce the density-density correlation function 
$S({\bf r},z)=\langle\hat\rho({\bf r},z)\hat\rho({\bf 0},0)\rangle$ in real space.
We have~:
\begin{eqnarray}
S({\bf r},z) & = & \int\frac{d^{d_\perp}\bf q}{(2\pi)^{d_\perp}}\int \frac{dq_z}{2\pi}
\;\;S({\bf q},q_z)\;\mbox{e}^{i{\bf q}\cdot{\bf r}}\mbox{e}^{iq_zz}
\end{eqnarray}
The integrations being quite straightforward for the free flux liquid considered here, we only quote the 
following intermediate result~:
\begin{eqnarray}
S({\bf q},z) = \rho\;\exp\Big(-\frac{k_BTq^2}{2\tilde\varepsilon_1}\;|z|\Big)
\end{eqnarray}
along with the final result
\begin{eqnarray}
S({\bf r}-{\bf r}',z-z') = \rho\;\Big(\frac{2\pi\tilde\varepsilon_1}{k_BT|z-z'|}\Big)^{d_\perp/2}
\;\exp\Big(-\frac{\tilde\varepsilon_1}{2k_BT}\,\frac{({\bf r}-{\bf r}')^2}{|z-z'|}\Big)
\end{eqnarray}
which shows that the density-density correlation function $\langle\hat\rho({\bf r},z)\hat\rho({\bf r},z')\rangle$
at the same transverse location ${\bf r}$ behaves like
\begin{eqnarray}
S({\bf 0},z) = \rho\;\Big(\frac{2\pi\tilde\varepsilon_1}{k_BT|z-z'|}\Big)^{d_\perp/2}
\end{eqnarray}
and thus we see that, even for the fully entangled noninteracting flux line liquid, density-density correlations 
$\langle\hat\rho({\bf r},z)\hat\rho({\bf r},z')\rangle$ at the same transverse location ${\bf r}$
decrease only algebraically as a function of the height separation $|z-z'|$. 
It is therefore very surprizing that $S({\bf 0},z)$ was found to decrease 
exponentially, like $\sim\exp(-|z|/\xi_v)$, 
in the presence of interactions in some numerical simulations of the uniformly frustrated 3D XY 
model\cite{Olsson-Teitel}.

\subsection{Weakly entangled flux liquid}

In a weakly entangled flux line liquid, $\langle u^2\rangle$ does not diverge with the sample thickness $L$, and 
therefore the first term on the {\em rhs} of equation (\ref{interm-3}) has to be kept. Using
equation (\ref{gen-conf-corr}),
\begin{eqnarray}
\langle[{\bf u}(z)-{\bf u}(z')]^2\rangle \simeq \frac{d_\perp k_BT a}{\sqrt{\pi}\varepsilon\varepsilon_0}
\big(1-\mbox{e}^{-\frac{\eta\sqrt\pi}{\varepsilon a}\,|z-z'|}\big)
\nonumber
\end{eqnarray}
we obtain~:
\begin{eqnarray}
\langle\hat\rho({\bf q},q_z)\hat\rho(-{\bf q},-q_z)\rangle & = &
L^2\delta_{q_z,0}\;L_\perp^{d_\perp}\,\rho^2g_0({\bf q})\;
\mbox{e}^{-\frac{1}{d_\perp}q^2\langle u^2\rangle}
+ NL\int dZ\;
\exp\Big(-q^2\,\frac{k_BT a}{2\sqrt{\pi}\varepsilon\varepsilon_0}
\big(1-\mbox{e}^{-\frac{\eta\sqrt\pi}{\varepsilon a}\,|Z|}\big)
\,\Big)
\;\mbox{e}^{-iq_zZ}
\nonumber\\
& = & LL_\perp^{d_\perp}\Big\{
L\delta_{q_z,0}\,\rho^2g_0({\bf q})\;
\mbox{e}^{-\frac{1}{d_\perp}q^2\langle u^2\rangle}
+ \rho\,\int dZ\;
\exp\Big(-q^2\;\frac{k_BT a}{2\sqrt{\pi}\varepsilon\varepsilon_0}
\big(1-\mbox{e}^{-\frac{\eta\sqrt\pi}{\varepsilon a}\,|Z|}\big)
\,\Big)
\;\mbox{e}^{-iq_zZ}
\Big\}\nonumber
\end{eqnarray}
from which we see that 
$S({\bf q},q_z)=\frac{1}{LL_\perp^{d_\perp}}\langle\hat\rho({\bf q},q_z)\hat\rho(-{\bf q},-q_z)\rangle$ 
is given by
\begin{eqnarray}
S({\bf q},q_z) = L\delta_{q_z,0}\,\rho^2g_0({\bf q})\;
\mbox{e}^{-q^2\frac{k_BT a}{2\eta\sqrt{\pi}\varepsilon\varepsilon_0}}
+ \rho\,\int dZ\;
\exp\Big(-q^2\;\frac{k_BT a}{2\sqrt{\pi}\varepsilon\varepsilon_0}
\big(1-\mbox{e}^{-\frac{\eta\sqrt\pi}{\varepsilon a}\,|Z|}\big)
\,\Big)
\;\mbox{e}^{-iq_zZ}
\end{eqnarray}
where we used the fact that
$\langle u^2\rangle \approx \big(d_\perp\,k_BT\,a/2\eta\sqrt{\pi}\varepsilon\varepsilon_0\big)$,
equation (\ref{gen-conf-u2}).
In view of the fact that the integration on the {\em rhs} of this last equation cannot be performed  
exactly, it is more convenient to use the partial Fourier transform~:
\begin{eqnarray}
S({\bf q},z) = \rho^2g_0({\bf q})\;
\mbox{e}^{-q^2 \frac{k_BT a}{2\eta\sqrt{\pi}\varepsilon\varepsilon_0}}
+ \rho
\exp\Big(-q^2\;\frac{k_BT a}{2\eta\sqrt{\pi}\varepsilon\varepsilon_0} 
\big(1-\mbox{e}^{-\frac{\eta\sqrt\pi}{\varepsilon a}\,|z|}\big)
\,\Big)
\label{interm-2-2}
\end{eqnarray}
which, with $\eta=1$, is the result (\ref{conf-struc-fac}) of section \ref{Theory-disentangled}.
On the other hand, for $\eta\ll 1$, the ``Debye-Waller'' factor 
$\exp\big(-q^2 \frac{k_BT a}{2\eta\sqrt{\pi}\varepsilon\varepsilon_0}\big)$
can be very small. Neglecting the first term on the {\em rhs} of 
equation (\ref{interm-2-2}), we obtain
\begin{eqnarray}
S({\bf q},z) \simeq
\rho
\exp\Big(-q^2\;\frac{k_BT a}{2\eta\sqrt{\pi}\varepsilon\varepsilon_0}
\big(1-\mbox{e}^{-\frac{\eta\sqrt\pi}{\varepsilon a}\,|z|}\big)
\,\Big)
\nonumber
\end{eqnarray}
which, for $|z|< \ell_z=\varepsilon a/\eta\sqrt\pi$ leads to equation (\ref{rho-rho-f}) of the text.

\begin{multicols}{2}

\end{multicols}

\begin{references}
\bibitem{Bednorz-Muller} J.G. Bednorz and K.A. Muller, Z. Phys. {\bf 64}, 189 (1986). 
\bibitem{Gammel1} P.L. Gammel, D.J. Bishop, G.J. Dolan, J.R. Kwo, C.A. Murray, L.F. Schneemeyer and J.V. Waszczak,
Phys. Rev. Lett. {\bf 59}, 2592 (1987).
\bibitem{Gammel2} P.L. Gammel, L.F. Schneemeyer, J.V. Waszczak and D.J. Bishop, Phys. Rev. Lett. {\bf 61}, 1666 
(1988).
\bibitem{Safar1} H. Safar, P.L. Gammel, D.A. Huse, D.J. Bishop, J.P. Rice and D.M. Ginsberg,
Phys. Rev. Lett. {\bf 69}, 824 (1992).
\bibitem{Safar2} H. Safar, P.L. Gammel, D.A. Huse, D.J. Bishop, W.C. Lee and D.M. Ginsberg,
Phys. Rev. Lett. {\bf 70}, 3800 (1993).
\bibitem{Charalambous} M. Charalambous, J. Chaussy, P. Lejay and V.M. Vinokur,
Phys. Rev. Lett. {\bf 71}, 436 (1993).
\bibitem{Kwok1} W.K. Kwok, J. Fendrich, S. Fleshler, U. Welp, J. Downe and G.W. Crabtree,
Phys. Rev. Lett. {\bf 72}, 1088 (1994).
\bibitem{Kwok2} W.K. Kwok, J. Fendrich, S. Fleshler, U. Welp, J. Downe and G.W. Crabtree,
Phys. Rev. Lett. {\bf 72}, 1092 (1994).
\bibitem{Cubitt} R.Cubitt, E.M. Forgan, G. Yang, S.L. Lee, D.M. Paul, H.A. Mook, M. Yethiraj, P.H. Kes, T.W. Li, 
A.A. Menovsky, Z. Tarnawski and K. Mortensen, Nature (London) {\bf 365}, 407 (1993).
\bibitem{Zeldov} E. Zeldov, D. Majer, M. Konczykowski, V.B. Geshkenbein, V.M. Vinokur and H. Shtrikman, Nature 
(London) {\bf 375}, 373 (1995). 
\bibitem{Abrikosov} A.A. Abrikosov, {\em Sov. Phys. JETP} {\bf 5}, 1174 (1957).
\bibitem{Nelson} D.R. Nelson, Phys. Rev. Lett. {\bf 60}, 1973 (1988).
\bibitem{Nelson-Seung} D.R. Nelson and H.S. Seung, Phys. Rev. B {\bf 39}, 9153 (1989).
\bibitem{Houghton-et-al} A. Houghton, R.A. Pelcovits and A. Sudb{\o}, Phys. Rev. B {\bf 42}, 906 (1990).
\bibitem{Ma-Chui} H.-R. Ma and S.T. Chui, Phys. Rev. Lett. {\bf 67}, 505 (1991).
\bibitem{Sengupta} S. Sengupta, C. Dasgupta, H.R. Krishnamurthy, G.I. Menon and T.V. Ramakrishnan, 
Phys. Rev. Lett. {\bf 67}, 3444 (1991).
\bibitem{Li-Teitel} Y.H. Li and S. Teitel, Phys. Rev. Lett. {\bf 66}, 3301 (1991); Phys. Rev. B {\bf 47}, 
359 (1993).
\bibitem{Ryu-et-al} S. Ryu, S. Doniach, G. Deutscher and A. Kapitulnik, 
Phys. Rev. Lett. {\bf 68}, 710 (1992).
\bibitem{Hetzel-et-al} R.E. Hetzel, A. Sudb{\o}, and D.A. Huse,
Phys. Rev. Lett. {\bf 69}, 518 (1992).
\bibitem{Dodgson-et-al} M.J.W. Dodgson, V.B. Geshkenbein, H. Nordborg and G. Blatter,
Phys. Rev. Lett. {\bf 80}, 837 (1998).
\bibitem{Blatter-et-al} G. Blatter, M.V.~Feigel'man, V.B. Geshkenbein, A.I. Larkin and V.M.~Vinokur, Rev. Mod.
Phys. {\bf 66}, 1125 (1994).
\bibitem{Fisher-Lee} M.P.A. Fisher and D.H. Lee, Phys. Rev. B {\bf 39}, 2756 (1989).
\bibitem{Negele-Orland} J.W. Negele and H. Orland, {\em Quantum Many Particle Systems}, Perseus Books, 1998.
\bibitem{Marchetti} C.M. Marchetti, Phys. Rev. B {\bf 43}, 8012 (1991).
\bibitem{Nelson2} D.R. Nelson, in {\em Phenomenology and Applications of High Temperature Superconductors},
edited by K.S.~Bedell et {\em al.}, Addison Wesley, 1992.
\bibitem{Feigelman-et-al} M.V. Feigel'man, V.B. Geshkenbein, L.B. Ioffe and A.I. Larkin,   
Phys. Rev. B {\bf 48}, 16 641 (1993).
\bibitem{Moore} M.A. Moore, Phys. Rev. B {\bf 55}, 14 136 (1997).
\bibitem{Hu-et-al} X. Hu, S. Miyashita and M. Tachiki, Phys. Rev. Lett. {\bf 79}, 3498 (1997); Phys. Rev. B {\bf
58}, 3438 (1998).
\bibitem{Nguyen-Sudbo-prb-98-2} A.K. Nguyen and A. Sudb{\o}, Phys. Rev. B {\bf 58}, 2802 (1998).
\bibitem{Righi-et-al} E.F. Righi, S.A. Grigera, G. Nieva, D. L\`opez and F. de la Cruz,
Phys. Rev. B {\bf 55}, 14 156 (1997).
\bibitem{Brandt} E.H. Brandt, J. Low Temp. Phys. {\bf 26}, 735 (1977); Physi\-ca C {\bf 165\& 166} 1129 (1990);
Int. J. Mod. Phys B {\bf 5}, 751 (1991); Physica {\bf C 195}, 1 (1992); Rep. Prog. Phys. {\bf 58}, 1465 (1995).
\bibitem{Barford-Gunn} W. Barford and J.M. Gunn, Physica {\bf C 156}, 515 (1988).
\bibitem{Sudbo-Brandt} A. Sudb{\o} and E.H. Brandt, Phys. Rev. B {\bf 43}, 10482 (1991).
\bibitem{Sardella} E. Sardella, Phys. Rev. B {\bf 45}, 3141 (1992).
\bibitem{remark0} See for instance Appendix B of reference\cite{Nelson-Vinokur}.
\bibitem{deGennes} P.G.~de Gennes, {\em Superconductivity of Metals and Alloys}, Addison-Wesley, 1966.
\bibitem{Marchetti-Nelson1} M.C. Marchetti and D.R. Nelson, Phys. Rev. B {\bf 42}, 9938 (1990).
\bibitem{Marchetti-Nelson2} M.C. Marchetti and D.R. Nelson, Physica C {\bf 174}, 40 (1991).
\bibitem{Radzihovsky-Frey} L. Radzihovsky and E. Frey, Phys. Rev. B {\bf 48}, 10357 (1993).
\bibitem{Tesanovic1} Z. Te{\v s}anovi\'c, Phys. Rev. B {\bf 51}, 16 204 (1995).
\bibitem{Tesanovic2} Z. Te{\v s}anovi\'c, Phys. Rev. B {\bf 59}, 6449 (1999).
\bibitem{Benetatos1} P. Benetatos and M.C. Marchetti, Phys. Rev. B {\bf 59}, 6499 (1999).
\bibitem{Benetatos2} P. Benetatos and M.C. Marchetti, preprint cond-mat/0101222 (January 2001).
\bibitem{Safar-et-al-prl-94} H. Safar, P.L. Gammel, D.A. Huse, S.N. Majumdar, L.F. Schneemeyer, D.J. Bishop, D.
L\'opez, G. Nieva and F. de la Cruz, Phys. Rev. Lett. {\bf 72}, 1272 (1994).
\bibitem{delaCruz-94} F. de la Cruz, D. L\'opez and G. Nieva, Philos. Mag. B {\bf 70}, 773 (1994).
\bibitem{Lopez-et-al-prl-96} D. L\`opez, E.F. Righi, G. Nieva and F. de la Cruz,
Phys. Rev. Lett. {\bf 76}, 4034 (1996).
\bibitem{Li-Teitel-prl-91} Y.H. Li and S. Teitel, Phys. Rev. Lett. {\bf 66}, 3301 (1991).
\bibitem{Li-Teitel-prb-93} Y.H. Li and S. Teitel, Phys. Rev. B {\bf 47}, 359 (1993).
\bibitem{Carneiro-et-al-prb-93} G. Carneiro, R. Cavalcanti and A. Gartner, Phys. Rev. B {\bf 47}, 5263 (1993).
\bibitem{Chen-Teitel-prl-94} T. Chen and S. Teitel, Phys. Rev. Lett. {\bf 72}, 2085 (1994).
\bibitem{Li-Teitel-prb-94} Y.H. Li and S. Teitel, Phys. Rev. B {\bf 49}, 4136 (1994).
\bibitem{Carneiro-prb94} G. Carneiro, Phys. Rev. B {\bf 50}, 6982 (1994).
\bibitem{Chen-Teitel-prl-95} T. Chen and S. Teitel, Phys. Rev. Lett. {\bf 74}, 2792 (1995).
\bibitem{Carneiro-prl-95} G. Carneiro, Phys. Rev. Lett. {\bf 75}, 521 (1995).
\bibitem{Nguyen-Sudbo-Hetzel-prl-96} A.K. Nguyen, A. Sudb{\o} and R.E. Hetzel, Phys. Rev. Lett. {\bf 77}, 1592
(1996).
\bibitem{Carneiro-prb-96} G. Carneiro, Phys. Rev. B {\bf 53}, 11 837 (1996).
\bibitem{Hagenaars-et-al} T.J. Hagenaars, E.H. Brandt, R.E. Hetzel, W. Hanke, M. Leghissa and G. Saemann-Ischenko,
Phys. Rev. B {\bf 55}, 11706 (1997).
\bibitem{Chen-Teitel-prb-97-1} T. Chen and S. Teitel, Phys. Rev. B {\bf 55}, 11 766 (1997).
\bibitem{Chen-Teitel-prb-97-2} T. Chen and S. Teitel, Phys. Rev. B {\bf 55}, 15 197 (1997).
\bibitem{Koshelev} A.E. Koshelev, Phys. Rev. B {\bf 56}, 11 201 (1997).
\bibitem{Ryu-Stroud-prl} S. Ryu and D. Stroud, Phys. Rev. Lett. {\bf 78}, 4629 (1997).
\bibitem{Ryu-Stroud} S. Ryu and D. Stroud, Phys. Rev. B {\bf 57}, 14 476 (1998).
\bibitem{Nordborg-Blatter-prl-97} H. Nodrborg and G. Blatter, Phys. Rev. Lett. {\bf 79}, 1925 (1997).
\bibitem{Nordborg-Blatter-prb-98} H. Nodrborg and G. Blatter, Phys. Rev. B {\bf 58}, 14 556 (1998).
\bibitem{Nguyen-Sudbo-prb-98-1} A.K. Nguyen and A. Sudb{\o}, Phys. Rev. B {\bf 57}, 3123 (1998).
\bibitem{Nguyen-Sudbo-prb-99} A.K. Nguyen and A. Sudb{\o}, Phys. Rev. B {\bf 60}, 15 307 (1999).
\bibitem{Olsson-Teitel} P. Olsson and S. Teitel, Phys. Rev. Lett. {\bf 82}, 2183 (1999).
\bibitem{Chin-et-al} S.-K. Chin, A.K. Nguyen and A. Sudb\o, Phys. Rev. B {\bf 59}, 14 017  (1999).
\bibitem{Doi-Edwards} See for example M. Doi and S.F. Edwards, {\em The Theory of Polymer Dynamics},
Oxford University Press, 1986.
\bibitem{McQuarrie} D.A. McQuarrie, {\em Statistical Mechanics}, Harper and Row, New York, 1976.
\bibitem{Hansen} J.P. Hansen and I.R. McDonald, {\em Theory of Simple Liquids}, Academic Press, London, 1986.
\bibitem{Feynman-Hibbs} R.P. Feynman and A.R. Hibbs, {\em Quantum Mechanics and Path Integrals}, McGraw-Hill,
New-York, 1965.
\bibitem{Fisher-et-al-73} M.E. Fisher, M.N. Barber and D. Jasnow, Phys. Rev. A {\bf 8}, 1111 (1973).
\bibitem{Ohta-79} T. Ohta and D. Jasnow, Phys. Rev. B {\bf 20}, 139 (1979). 
\bibitem{Miyashita-84} S. Miyashita and H. Shiba, J. Phys. Soc. Jpn. {\bf 53}, 1145 (1984).
\bibitem{Ceperley} D.M. Ceperley, Rev. Mod. Phys. {\bf 67}, 279 (1995); and references therein.
\bibitem{Glazman-Koshelev} L.I. Glazman and A.E. Koshelev, Phys. Rev. B {\bf 43}, 2835 (1991). 
\bibitem{Remark3} To avoid any misunderstanding, here we are talking about equation (24) of reference
\cite{Glazman-Koshelev}.
\bibitem{Frey-et-al-prb-94} Another scenario for the destruction of longitudinal phase coherence has
been proposed by E. Frey, D.R. Nelson and D. Fisher, Phys. Rev. B {\bf 49}, 9723 (1994).
\bibitem{Klemm-et-al-75} R.A. Klemm, A. Luther and M.R. Beasley, Phys. Rev. B {\bf 12}, 877 (1975).
\bibitem{Remark2} For more details, see Appendix B of ref.\cite{Nordborg-Blatter-prb-98}.
\bibitem{Nelson-Vinokur} D.R. Nelson and V.M. Vinokur, Phys. Rev. B {\bf 48}, 13060 (1993).
\bibitem{Prudnikov} A.P. Prudnikov, Yu.A. Brychkov and O.I.~Marichev, {\em Integrals and Series},
Gordon and Breach, New York, 1986.
\bibitem{Abramowitz} {\em Handbook of Mathematical Functions}, M.~Abramo\-witz and I.A.~Stegun (Editors), 
Dover, 1965.
\end{references}
\end{document}